\documentclass[11pt]{article}
\pdfoutput=1

\usepackage[english]{babel} 
\usepackage{scalerel}
\usepackage{stackrel}
\usepackage{amssymb,amsmath,amsthm,mathrsfs,latexsym,amsxtra,graphicx,appendix,epstopdf,feynmf,hyperref,setspace,fix-cm,color,cite,hyperref}
\usepackage{verbatim}
\usepackage[a4paper,left=2cm,right=1cm,top=4cm,bottom=4cm,bindingoffset=5mm]{geometry}
\usepackage{pgfplots}
\pgfplotsset{compat=newest}
\usepackage{mathtools}
\usepackage{tensor}
\usepackage{tikz}
\usetikzlibrary{trees}
\usetikzlibrary{decorations.pathmorphing}
\usetikzlibrary{decorations.markings}
\usetikzlibrary{decorations.markings}
\usepackage{float}
\usepackage{rotating}
\usepackage{simplewick}
\usepackage{cancel}
\usepackage{ marvosym }
\usepgfplotslibrary{fillbetween}
\usetikzlibrary{patterns}
\usepackage{physics}

\usepackage{makecell}

\usepackage{mathtools}

\usepackage{setspace}
 \usepackage{multirow}
\usepackage{fullpage}
\usepackage{amsmath}
\usepackage{amssymb}
\usepackage{setspace}
\usepackage{bbm}
\usepackage{dsfont}
\usepackage{graphics}
\usepackage{subcaption}
\usepackage{longtable}
\usepackage[font=footnotesize,labelfont=bf,width=.9\textwidth]{caption}
\usepackage{color}
\usepackage{etoolbox}
 \usepackage{multirow}
\usepackage{booktabs}
\usepackage{array}
\usepackage{hyperref}
\usepackage{cite}
\usepackage[normalem]{ulem}
\hypersetup{
    colorlinks,%
    citecolor=blue,%
    filecolor=blue,%
    linkcolor=blue,%
    urlcolor=blue
}
\usepackage{bm}
\allowdisplaybreaks
\def\be{\begin{equation}}
\def\ee{\end{equation}}
\def\bea{\begin{align}}
\def\eea{\end{align}}
\newcommand{\beq}{\begin{eqnarray}}
\newcommand{\eeq}{\end{eqnarray}}

\renewcommand{\dd}{\mathrm{d}}

\newcommand{\p}{\nabla}

\newcommand{\lp}{\left(}
\newcommand{\rp}{\right)}
\newcommand{\Y}{\mathcal{Y}}

\def\({\left(}
\def\){\right)}
\def\[{\left[}
\def\]{\right]}

\def\p{\partial}

\def\le{\left}
\def\ri{\right}

\newcommand{\bpz}{\bm{\psi}}
\newcommand{\bph}{\bm{\varphi}}
\newcommand{\bch}{\bm{\chi}}

\newcommand{\barc}{\bar{\chi}}
\newcommand{\barp}{\bar{\varphi}}

\renewcommand{\title}[1]{\vbox{\center\LARGE{#1}}\vspace{5mm}}
\renewcommand{\author}[1]{\vbox{\center#1}\vspace{5mm}}
\newcommand{\address}[1]{\vbox{\center\footnotesize\em#1}}
\newcommand{\email}[1]{\vbox{\center\footnotesize\tt#1}\vspace{5mm}}

\begin{document}
\numberwithin{equation}{section}
{
\begin{titlepage}

\begin{center}

\hfill \\
\hfill \\
\vskip 1cm

\title{\bf Near-AdS$_2$ Spectroscopy: classifying the spectrum of operators and interactions in $\mathcal{N} = 2$ 4D supergravity
}


\author{Alejandra Castro and  Evita Verheijden
}
\address{
 Institute for Theoretical Physics, University of Amsterdam, 
1090 GL Amsterdam, The Netherlands
}

\email{ a.castro@uva.nl, e.m.h.verheijden@uva.nl  }

\end{center}

\vskip 1cm

\begin{center} {\bf ABSTRACT } \end{center}
We describe holographic properties of near-AdS$_2$ spacetimes that arise within spherically symmetric configurations of ${\cal N}=2$ 4D $U(1)^4$ supergravity, for both gauged and ungauged theories. These theories pose a rich space of AdS$_2\times S^2$ backgrounds, and their responses in the near-AdS$_2$ region are not universal. In particular, we show that the spectrum of operators dual to the matter fields, and their cubic interactions, are sensitive to properties of the background and the theory it is embedded in. The properties that have the most striking effect are whether the background is supersymmetric or not, and if the theory is gauged or ungauged. Interesting effects are due to the appearance of operators with $\Delta < 2$, which depending on the background can lead to, for instance, instabilities or extremal correlators. The resulting differences will have an imprint on the quantum nature of the microstates of near-extremal black holes, reflecting that not all extremal black holes respond equally when kicked away from extremality.  
\vfill

\end{titlepage}
}


\newpage

{
\baselineskip12pt
  \hypersetup{linkcolor=black}
  \tableofcontents
}

\newpage 
 \section{Introduction}

Recent years have seen great progress in understanding the quantum properties of black holes in the context of the AdS/CFT correspondence. While the universal nature of the Bekenstein-Hawking area-law reflects that there should be a commonality in the statistical origin of their entropy, this might not be the case for a refined description. 
Our goal here is to underscore aspects of black holes that are not universal, despite their shared semi-classical features. In particular we will bring to light concise holographic data that are sensitive to the surrounding theory that fosters the black hole and to the interplay of the theory content with properties of the background. 
 
Our analysis is embedded in the developments coined as the near-AdS$_2$/near-CFT$_1$ correspondence \cite{Almheiri:2014cka,Maldacena:2016upp}. This instance of holography describes deformations away from an idealized AdS$_2$ geometry, which are relevant to construct a holographic description of the near-horizon region of near-extremal black holes. One of the most prominent results of near-AdS$_2$/near-CFT$_1$ was to show that the leading gravitational backreaction, which defines near-AdS$_2$, is  universally encoded in two-dimensional Jackiw-Teitelboim (JT) gravity \cite{Jackiw:1984je,Teitelboim:1983ux}. This is the commonality that is nowadays used to decode quantum properties of black holes. Building on these developments, here  we wish to further decode what other degrees of freedom appear as one backreacts the geometry, and how these degrees of freedom interact with the JT sector. In particular we will report on the spectrum of fluctuations in the near-horizon region, which we translate to the spectrum of operators in the CFT$_1$, and cubic interactions among these fluctuations.   
 
We will focus on some of the simplest scenarios of near-AdS$_2$ that one might expect to lead to a UV-complete description in string theory: backgrounds that arise as solutions to a four-dimensional supergravity theory and connect to suitable near-extremal black holes therein. With regard to these theories, we will study solutions of four-dimensional ${\cal N}=2$ supergravity, covering both gauged and ungauged cases which we describe in more detail below. The overarching feature of these theories is that they contain four $U(1)$ gauge fields and six real scalar fields (with three of them being axions). We will consider configurations in four dimensions that respect spherical symmetry, where the extremal near-horizon region is precisely AdS$_2\times S^2$. This makes it possible to build an effective two-dimensional description by integrating out the 2-sphere, and tie our results to the features of JT gravity in a simple manner. 

The strategy presented here is as follows. We will start with an AdS$_2\times S^2$ background that solves the equations of four-dimensional ${\cal N}=2$ supergravity. We will first proceed to study linearized perturbations around this background that preserve spherical symmetry. From here we will single out the JT sector, which encodes the features of near-AdS$_2$; we will also organize the remaining perturbations according to their scaling dimension, and interpret them as dual operators. The second step is to characterize the cubic interactions among these operators and the JT sector, and quantify the leading correction of their two-point functions due to these interactions. This closely follows the analysis in \cite{Castro:2021fhc}, and is also mentioned in \cite{Maldacena:2016upp,Narayan:2020pyj}.

There are two important features that have the biggest imprint on our analysis. The first one is the  {\it ``background''}: our starting point is an AdS$_2$ solution, characterized by a set of electric and magnetic charges, and the  six scalar fields are controlled by the attractor mechanism \cite{Ferrara:1995ih,Strominger:1996kf,Ferrara:1996dd,Ferrara:1997tw,Goldstein:2005hq,Alishahiha:2006ke,Kallosh:2006bt,Ceresole:2007wx}. As we consider different AdS$_2$ backgrounds in the gauged and ungauged theory, the most important feature of each background is if the solution preserves supersymmetry, i.e.\ if the extremal black hole is BPS or non-BPS.\footnote{BPS stands for saturation of the Bogomolnyi-Prasad-Sommerfield bound, which here we use to denote that the solution preserves a fraction of supersymmetry. Non-BPS denotes that the solution preserves no supersymmetry.} The spectrum of operators is highly sensitive to this feature. In the ungauged theory, the non-BPS black holes will contain marginal operators, which is a direct consequence of having a flat direction in the attractor mechanism \cite{Tripathy:2005qp}; these operators will lead to extremal three-point correlators that are pathological. On the other hand, the operators in the BPS branch in the ungauged theory all have conformal dimension $\Delta=2$, placing them on similar footing to the JT sector, in agreement with the nAttractor proposed in \cite{Larsen:2018iou}. For the gauged theory, non-BPS AdS$_2$ backgrounds have modes that violate the Breitenlohner-Freedman (BF) stability bound in AdS$_2$ \cite{Breitenlohner:1982bm}, which signals an instability along the lines of \cite{Gubser:2008px}. BPS black holes, in contrast, do not exhibit these instabilities, nor do they contain marginal operators. 

The second crucial feature comes from the  {\it ``surrounding''}: the theory within which the solution is embedded. Here it is important to account for all scalar fields present in the truncation, and if the theory is gauged or ungauged. The presence of the cosmological constant in the gauged theory, in comparison to the ungauged theory, tends to lower the conformal dimensions of the operators. For non-BPS backgrounds, this leads to the BF instabilities aforementioned, which are unique to the gauged theory. For BPS backgrounds, the gauged theory will contain relevant operators, and irrelevant ones that have $1<\Delta<3/2$; the presence of this matter content indicates that the Schwarzian mode of JT gravity is not supposed to be the dominant effect (see \cite{Maldacena:2016upp}).  
The behaviour of the interactions is also rather different: for the backgrounds in the gauged theory the cubic couplings allow for both positive and negative signs, depending on the conformal dimensions of the fields, while in the ungauged theory they have definite signs.

 
 Although we are restricting the discussion here to AdS$_2\times S^2$ near-horizon backgrounds, relevant for static dyonic black holes, we will only discuss certain solutions. Their properties and relation to extremal black holes are the following.
 \begin{itemize}
     \item {\bf BPS branch, ungauged theory:} our analysis covers the most general dyonic solution with four magnetic and four electric charges. We will follow conventions of \cite{Chow:2014cca}, where the corresponding black hole solution is described in detail. A more recent discussion on extremal dyonic black holes can be found in \cite{Cvetic:2021lss}.
     \item {\bf Non-BPS branch, ungauged theory:} we will cover a large class of non-BPS solutions, but with certain limitations since the attractor equations generically admit non-linear solutions. The corresponding black hole solution will again follow \cite{Chow:2014cca}, and prior works of interest here include  \cite{Goldstein:2005hq,GimonLarsenSimon2008,Bellucci:2008sv} and references therein.
     \item {\bf Magnetic, BPS, gauged theory:} these are backgrounds that carry four magnetic charges, and preserve supersymmetry. The corresponding black holes were first constructed in \cite{Cacciatori:2009iz}, and accounting for the Bekenstein-Hawking entropy in AdS$_4$/CFT$_3$ was first done in \cite{Benini:2015eyy}.
     \item {\bf Magnetic, non-BPS, gauged theory:} we focus on magnetic solutions that smoothly interpolate from the gauged to the ungauged theory, and hence are not supersymmetric within AdS$_4$. These black holes date back to \cite{Cvetic:1999xp}, and our conventions follow \cite{Chow:2013gba}.
    \item {\bf Dyonic, non-BPS, gauged theory:} to illustrate the effects of dyonic backgrounds in the gauged theory, we consider a simple case with only one electric and one magnetic charge. The corresponding black hole is in  \cite{Chow:2013gba}.
 \end{itemize}
 As reflected by this list, static black holes embedded in AdS$_4$ are notoriously more difficult to construct and analyse, which is the reason that our examples in the gauged theory are more limited.  
 Other solutions that we are not considering here, but are worth investigating, are the most general dyonic static BPS black holes in AdS$_4$ gauged supergravity  \cite{Katmadas:2014faa,Halmagyi:2014qza} --- with their statistical interpretation via the dual CFT done in \cite{Benini:2016rke}. There are as well many other non-BPS solutions in the gauged theory, see e.g. \cite{Gnecchi:2012kb,Chow:2013gba}, and one could also consider near-horizon geometries of the form  AdS$_2\times \Sigma_g$, with $\Sigma_g$ a two-dimensional Riemann surface of genus $g$, which we will not do here. 
 
 In the context of  near-AdS$_2$/near-CFT$_1$, and its ties to near-extremal black holes, the Reissner-Nordstr\"{o}m solution was  an important lamppost in these developments. Some of the original references are \cite{Almheiri:2016fws,Nayak:2018qej}, which also include dyonic cases. An important generalization here is to embed these solutions carefully within a supergravity theory, and quantify the effects that the background and surrounding have on the dictionary that dictates properties of the near-CFT$_1$.\footnote{At extremality, a recent analysis of AdS$_2\times S^2$ solutions of ${\cal N}=2$ ungauged supergravity viewed from the perspective of the 2D gravity can be found in \cite{Aniceto:2020saj}, which includes higher derivative corrections. An excellent review, using the technology of the quantum entropy function, is presented in \cite{Sen:2005wa}.} These properties, the spectrum of operators and interactions, show to us how building a statistical interpretation of the different cases explored here is already highly non-trivial, and it will require a more intricate dual description --- in comparison to the SYK-like models studied in, for instance, \cite{Gross:2016kjj,Anninos:2020cwo,Fu:2016vas,Murugan:2017eto,Marcus:2018tsr}.

This paper is organized as follows. We start in Sec.\,\ref{sec:sec2} by introducing the four-dimensional $\mathcal{N} = 2$ supergravity theory and its field content, and the dimensional reduction to AdS$_2{ \times S^2}$. In Sec.\,\ref{sec:generalaspects} we describe general aspects of the near-AdS$_2$ analysis. We discuss the AdS$_2$ background and the attractor mechanism, and study linearized perturbations of the dilaton, scalar fields and the metric around this background. We single out the JT sector, and discuss the cubic interactions of the matter fields with this sector. We quantify the correction of these interactions on the two-point functions of the matter fields. In Sec.\,\ref{sec:ungauged} we evaluate these expressions explicitly for the ungauged theory. We classify the solutions according to them being BPS or non-BPS, and contrast the responses of the near-AdS$_2$ region and the corrections to the two-point functions in both branches. As mentioned above, we find that fluctuations around BPS backgrounds comply with the nAttractor mechanism, but for non-BPS solutions we find a marginal operator corresponding to a flat direction in the attractor mechanism. For non-BPS backgrounds we find non-vanishing cubic couplings for certain extremal correlators, and discuss the repercussions of this pathology.
In Sec.\,\ref{sec:gauged-cases} we consider some special cases in the gauged theory: purely magnetic solutions, both BPS and non-BPS, and a dyonic non-BPS solution with a single charge. We classify the non-universal features at the level of the spectrum of operators and the interactions. We end in Sec.\,\ref{sec:disc} with a careful discussion of our results, and discuss future directions. 
We also included four appendices: in App.\,\ref{app:conventions} we gather some basic conventions and notation; in App.\,\ref{app:Fterms} we collect explicit formulas necessary to integrate out the field strengths for our $U(1)^4$ supergravity theory, and the attractor solutions for this theory (for general charges and purely magnetic solutions). In App.\,\ref{app:magnetic} we give explicit expressions for the linearized equations and interactions for purely magnetic (BPS and non-BPS) backgrounds. Finally, in App.\,\ref{app:scalings} we explain how to set up the extremal limit for both BPS and non-BPS black hole solutions, and give a numerical example illustrating the appearance of a flat direction in the linear response.

\section{Two-dimensional effective field theory description} \label{sec:sec2}
 
\subsection{\texorpdfstring{${\cal N}=2$ $U(1)^4$}{N=2 U(1)4} gauged supergravity}\label{sec:gaugedsugra}
 
Our analysis is centered around bosonic solutions to ${\cal N}=2$ $U(1)^4$ gauged supergravity in four dimensions. This theory can be viewed as an Abelian truncation to the Cartan subgroup $U(1)^4$ of ${\cal N}=8$ $SO(8)$ gauged supergravity \cite{Cvetic:1999xp}. Our conventions follow those in \cite{Chow:2013gba}, which we quickly summarize here. Note that these conventions differ in very minor ways relative to e.g.\ \cite{Cvetic:1999xp,Chong:2004na}, where the definitions of the scalar fields are slightly different relative to the description here.
 
The basic ingredients of this  supergravity theory are the following. The bosonic fields are the metric $g^{(4)}_{\mu\nu}$, six real scalar fields, and four gauge fields. The scalar fields are split into three dilatons, $\varphi_i$, and three axions, $\chi_i$ with $i=1,2,3$; the gauge fields are labeled as $A^I$ with $I=1,2,3,4$.   Because we are in four dimensions, there are several formulations of the action since the gauge fields can be dualized; the dual gauge field will be denoted as $\tilde A_I$, and the corresponding field strengths are
 \be\label{eq:defF}
 F^I = dA^I ~,\qquad \tilde{F}_I = d\tilde A_I~.
 \ee
To describe the dynamical aspects of the theory, we will use here the dual formulation as described in \cite{Chow:2013gba}, where the action is given by
\be\label{eq:action4d}
I_{\rm 4D}= \frac{1}{16\pi G_4} \int d^4x \sqrt{-g^{(4)}}{\cal L}_{\rm 4D} ~,
\ee
and
 \be
 \begin{aligned}\label{eq:L4D}
{\cal L}_{\rm 4D} = R^{(4)} &-\frac{1}{2}\sum_{i=1}^3 \le( (\partial \varphi_i)^2  + e^{2\varphi_i} (\partial \chi_i )^2\ri) 
+g^2\sum_{i=1}^3\le(2\cosh\varphi_i+\chi_i^2e^{\varphi_i}\ri) \\
& -\frac{1}{4}e^{-\varphi_1}\le(e^{\varphi_2+\varphi_3} ({\cal F}^1)^2+e^{\varphi_2-\varphi_3} (\tilde{\cal F}_2)^2
+ e^{-\varphi_2+\varphi_3}(\tilde{\cal F}_3)^2 +e^{-\varphi_2-\varphi_3}( {\cal F}^4)^2 \ri)\\
& -\frac{1}{4}\chi_1\,  \epsilon^{\mu\nu\alpha\beta}\left(F^1_{\mu\nu}F^4_{\alpha\beta} + \tilde F_{2\mu\nu} \tilde F_{3\alpha\beta}\right)~.
 \end{aligned}
 \ee
Here $g$ is a real gauge-coupling constant, which effectively acts as a negative cosmological constant; setting $g=0$ gives rise to the STU model in ungauged supergravity. The calligraphic field strengths are related to those in \eqref{eq:defF} via
 \be\label{eq:dicFcF}
 \begin{aligned}
 {\cal F}^1&=F^1+\chi_3 \tilde F_2 +\chi_2 \tilde F_3-\chi_2\chi_3 F^4~,\cr
 \tilde {\cal F}_2&=\tilde F_2-\chi_2  F^4~, \cr
 \tilde {\cal F}_3&=\tilde F_3-\chi_3  F^4 ~,\cr
  {\cal F}^4&=F^4~,
 \end{aligned}
 \ee
where two of our gauge fields $(\tilde F_2,\tilde F_3)$ are treated in terms of their duals. It will be useful to rewrite the terms involving field strengths in \eqref{eq:L4D} as
 \be
 \begin{aligned}\label{eq:maxwell-action-mod}
  {\cal L}_{4D} =R^{(4)} &- \frac{1}{2}\sum_{i=1}^3 \le( (\partial \varphi_i)^2  + e^{2\varphi_i} (\partial \chi_i )^2\ri) 
+g^2\sum_{i=1}^3\le(2\cosh\varphi_i+\chi_i^2e^{\varphi_i}\ri) \\
&  -\frac{1}{4} k_{IJ}{\bf F}^I_{~\mu\nu} {\bf F}^{J\mu\nu}+\frac{1}{4} h_{IJ}  \epsilon^{\mu\nu\alpha\beta} {\bf F}^I_{~\mu\nu}{\bf F}^{J}_{\alpha\beta}~.
 \end{aligned}
 \ee
  We have introduced some notation to reflect that we are in a mixed situation where some fields are dualized: in our case of interest, we have
  \be
  {\bf F}^I \equiv (F^1,\tilde F_2, \tilde F_3, F^4)~.
  \ee
Expressions for the matrices $k_{IJ}$ and $h_{IJ}$ in terms of $\varphi_i$ and $\chi_i$ are written in App.\,\ref{app:Fterms}. 

 \subsection{Dimensional reduction}
 
In the following, we will perform the dimensional reduction of bosonic solutions to ${\cal N}=2$ $U(1)^4$ gauged supergravity that preserve spherical symmetry. The outcome of this subsection is an effective two-dimensional action that fully captures the dynamics of four-dimensional spherically symmetric backgrounds. There are several references that perform a very similar analysis to ours. We will mostly follow \cite{Larsen:2018iou} for the treatment of the gauge fields, and the metric will be treated as in, e.g., \cite{Nayak:2018qej}.

We will write backgrounds that preserve spherical symmetry as
\be\label{eq:4d2d}
ds^4 = \frac{1}{\Phi(x)} g_{ab} \dd x^a \dd x^b + \Phi^2(x) \le(\dd \theta^2+\sin^2\theta\dd \phi^2\ri)~.
\ee
Here $x^a$ are two-dimensional coordinates, with $a,b=0,1$,  and $g_{ab}$ is the two dimensional metric.\footnote{In the following, $x$ will be a shorthand referring to the two-dimensional coordinates $x^a$, and not the four-dimensional coordinates $x^\mu$. For example, $\Phi(x)\equiv \Phi(x^a)$.} The scalar $\Phi(x)$ is the radius of the 2-sphere, and it is also introduced as a conformal factor for $g_{ab}$. The relative powers of $\Phi$ in \eqref{eq:4d2d} are selected such that the four-dimensional Ricci scalar is related to its two-dimensional counterpart as
	\begin{equation}\label{eq:ricci42}
		R^{(4)} = \Phi R^{(2)} + \frac{2}{\Phi^2} - \frac{3}{\Phi} g^{ab} \nabla_a(\Phi \nabla_b \Phi)~.
	\end{equation}
The last term here will correspond to a total derivative when replaced back in  \eqref{eq:action4d}, and hence there will be no explicit kinetic terms for $\Phi(x)$ in our final answer; $\Phi(x)$  will be referred to as the dilaton. In the following, we will discuss the remaining matter fields --- four field strengths and six scalars --- and write the resulting two-dimensional theory obtained by placing \eqref{eq:action4d} on the background \eqref{eq:4d2d}. We are also only focusing on the so-called $s$-wave sector of the theory, which implies that the matter fields will respect the isometries of the 2-sphere made manifest in \eqref{eq:4d2d}. For the six scalars $(\varphi_i,\chi_i)$, this means that we will assume they only depend on $x^a$.

Next, we focus on the field strengths supported by the background \eqref{eq:4d2d}. 
Because of the spherical symmetry, and since the spacetime is a direct product, one can solve in full generality for the field strengths that respect this structure, i.e.\ one can integrate them out. To be concrete, the general two-form that conforms with the symmetries of \eqref{eq:4d2d}  will be of the form  
\be\label{eq:magF}
{F}^I = (\cdots) \epsilon_{ab}\dd x^a\wedge \dd x^b- {P}^I \sin\theta \dd \theta\wedge \dd\phi~,
\ee
and 
\be\label{eq:magtF}
\tilde F_I = (\cdots) \epsilon_{ab}\dd x^a\wedge \dd x^b+ Q_I \sin\theta \dd \theta\wedge \dd\phi~.
\ee
Here $\epsilon_{ab}$ is the epsilon tensor for $g_{ab}$. Note that $\sin\theta \dd \theta\wedge \dd\phi$ is the volume form in $S^2$, so this just reflects that $F^I$ and $F_I$ are linear combinations of the volume (top) forms on $g_{ab}$ and $S^2$. $P^I$ and $Q_I$ are constants that correspond to magnetic and  electric charges: notice that $F^I$ carries the magnetic charge in the angular components, while $\tilde F_I$ carries the electric charge in the angular components.\footnote{Our definition of $P^I$ and $Q_I$ is exactly the same as in, e.g., \cite{Chow:2013gba}, and they are presented  in geometrical units. Other suitable ways to normalize the charges are $\bar {P}^I = \frac{{P}^I}{4G_4}$ and $\bar {Q}_I = \frac{{Q}_I}{4G_4}$, or to introduce quantized charges as $p^I = \frac{{P}^I}{\sqrt{4G_4}}$ and $q_I = \frac{{Q}_I}{\sqrt{4G_4}}$.}  The $(\cdots)$ will be polynomials of the scalars $\Phi$, $\varphi_i$ and $\chi_i$ to be determined by imposing the equations of motion for the field strengths. This is what we want to write explicitly in the following. 

The equation of motion from varying with respect to ${\bf F}^I$ is given by
\be\label{eq:eomF}
\nabla_\mu\le( k_{IJ}{\bf F}^{J\,\mu\nu}- h_{IJ}  \epsilon^{\mu\nu\alpha\beta} {\bf F}^{J}_{\,\alpha\beta}\ri)=0~,
\ee
where we are adopting the notation in \eqref{eq:maxwell-action-mod}. Since we are assuming that $\varphi_i= \varphi_i(x)$ and $\chi_i=\chi_i (x)$, the angular components of this equation are automatically satisfied by \eqref{eq:magF}-\eqref{eq:magtF}. The components along $x^a$ have a simple solution given by
\be\label{eq:electF}
\Phi^3 k_{IJ}{\bf F}^{J}_{~ab} -2 h_{IJ} {\bf P}^J  \epsilon_{ab}  = {\bf Q}_I \epsilon_{ab}~,
\ee
where we have made use of \eqref{eq:4d2d} and cast the solution in terms of the two-dimensional metric $g_{ab}$. It is straightforward to invert this equation since both $k_{IJ}$ and $h_{IJ}$ have an inverse, and the explicit solution  is given in \eqref{eq:solcF}. We also introduced the bold notation for the charges:
  \be
  {\bf Q}_I \equiv (Q_1,P^2, P^3,Q_4)~,\qquad {\bf P}^I \equiv (P^1,-Q_2,-Q_3,P^4)~.
  \ee
With this notation we can simply rewrite \eqref{eq:magF}-\eqref{eq:magtF} as  
\be\label{eq:finalF}
{\bf F}^I= \frac{1}{2}{\bf F}^I_{ab} \dd x^a\wedge \dd x^b- {\bf P}^I \sin\theta \dd \theta\wedge \dd\phi~,
\ee
with ${\bf F}^I_{ab} $ determined by \eqref{eq:electF}. 

Given that it is very simple to solve for ${\bf F}^{J}$, in the process of constructing our effective two-dimensional theory, we will integrate them out; i.e.\ we want to trade ${\bf F}^I$ for $({\bf P}^I ,{\bf Q}_I)$. For the components of ${\bf F}^I$ along the 2-sphere this is a simple replacement of \eqref{eq:finalF} in the action. For ${\bf F}^I_{ab}$ this requires performing a Legendre transform to consistently trade it for ${\bf Q}_I$ in the action, and comply with the equations of motion; for a more detailed discussion see, for example, \cite{Larsen:2018iou}. The steps  are the following: start with the contribution to the action from the second line of \eqref{eq:maxwell-action-mod}, and replace \eqref{eq:4d2d} and \eqref{eq:finalF}, which gives
 \be\label{eq:steps1}
 \begin{aligned}
&\frac{1}{16\pi G_4} \int d^4x \sqrt{-g^{(4)}}\le(-\frac{1}{4} k_{IJ}{\bf F}^I_{~\mu\nu} {\bf F}^{J\mu\nu} + \frac{1}{4} h_{IJ}  \epsilon^{\mu\nu\alpha\beta} {\bf F}^I_{~\mu\nu}{\bf F}^{J}_{\alpha\beta}\ri)\\
&= \frac{1}{4 G_4}\int d^2 x \sqrt{-g^{(2)}} \Phi^3 \left( - \frac{1}{4}k_{IJ} ({\bf F}^I_{~ab} {\bf F}^{J\,ab}  +\frac{2}{\Phi^{6}} {\bf P}^I{\bf P}^J ) - \frac{h_{IJ}}{\Phi^3} \epsilon^{ab} {\bf F}^I_{~ab} {\bf P}^J  \right)~.
 \end{aligned}
 \ee
 The second step is the Legendre transform, which amounts to adding to the action the term
 	\begin{equation} \label{eq:steps2}
	- \frac{1}{8G_4} \int d^2x \sqrt{-g^{(2)}} \, \epsilon^{ab}{\bf Q}_I {\bf F}^I_{~ab}~.
	\end{equation}
	After adding this term to \eqref{eq:steps1} it is consistent to replace ${\bf F}^I_{~ab}$ by $({\bf P}^I ,{\bf Q}_I)$ via \eqref{eq:electF}.
	
Finally, incorporating all of our ingredients, we can write the effective two-dimensional theory. Using \eqref{eq:4d2d}, \eqref{eq:ricci42}, \eqref{eq:steps1}, and \eqref{eq:steps2}, our two-dimensional action and Lagrangian are
 \be\label{eq:steps3a}
 \begin{aligned}
 I_{\rm 2D} =  \frac{1}{4 G_4}\int d^2 x \sqrt{-g^{(2)}} \,{\cal L}_{\rm 2D}~,
  \end{aligned}
 \ee
 and
  \be\label{eq:steps3}
 \begin{aligned}
 {\cal L}_{\rm 2D}=&\,\Phi^2 R^{(2)} + \frac{2}{\Phi} +g^2\Phi\sum_{i=1}^3\le(2\cosh\varphi_i+\chi_i^2e^{\varphi_i}\ri)\\
&-\frac{\Phi^2}{2}\sum_{i=1}^3 \le( (\partial_a \varphi_i)(\partial^a \varphi_i)  + e^{2\varphi_i} (\partial^a \chi_i )(\partial_a \chi_i)\ri)- \frac{1}{2\Phi^3} V({\bf P},{\bf Q})~.
\end{aligned}
 \ee
Here, $V({\bf P},{\bf Q})$ is a scalar potential encoding the magnetic and electric charges,\footnote{$V({\bf P},{\bf Q})$ has several simplifications and identities that apply for the $U(1)^4$ theory. These are listed in App.\,\ref{app:Fterms}.}
 \be 	 \label{eq:defV}
V({\bf P},{\bf Q})\equiv ({\bf P}^I ~~  {\bf Q}_I)
   \begin{pmatrix}
     k_{IJ} + 4 h_{IK} (k^{-1})^{KL}h_{LJ} & -2 h_{IK}(k^{-1})^{KJ}\\
   -2 (k^{-1})^{IK}h_{KJ} & (k^{-1})^{IJ}\\
 \end{pmatrix}
   \begin{pmatrix}
 {\bf P}^J  \\
 {\bf Q}_J
 \end{pmatrix} ~.
 \ee
This Lagrangian is a consistent truncation for the $s$-wave sector of $U(1)^4$ gauged supergravity when compactified on $S^2$, and it will be the main object that we will use in the coming section.
 
Lastly, it will be useful to record the equations of motion associated to \eqref{eq:steps3}. Varying with respect to the dilaton gives
	\begin{equation}\label{eq:eomdilaton}
	\begin{aligned}
		 \Phi R^{(2)}  =& \, \frac{1}{\Phi^2}  - \frac{g^2}{ 2}\sum_{i=1}^3\le(2\cosh\varphi_i+\chi_i^2e^{\varphi_i}\ri) - \frac{3}{4\Phi^4} V\\
		  &+\frac{\Phi}{2} \sum_{i=1}^3 \le( (\partial_a \varphi_i)(\partial^a \varphi_i)  + e^{2\varphi_i} (\partial^a \chi_i )(\partial_a \chi_i)\ri) ~.
	 \end{aligned}
	\end{equation}
The equations of motion from varying the scalars $\varphi_i$ and $\chi_i$ are, respectively,
	\begin{equation}\label{eq:eomscalars}
	\begin{aligned}
		 \nabla_a \left( \Phi^2   \nabla^a \varphi_i \right) - \Phi^2 e^{2\varphi_i} g^{ab} \partial_a \chi_i \partial_b \chi_i + g^2 \Phi (2\sinh \varphi_i + \chi_i^2 e^{\varphi_i}) - \frac{1}{2\Phi^3} \partial_{\varphi_i} V = 0~,\\
 \nabla_a \left( \Phi^2  e^{2\varphi_i} \nabla^a \chi_i \right)  + 2g^2 \Phi\, \chi_i e^{\varphi_i} - \frac{1}{2\Phi^3} \partial_{\chi_i} V = 0~.	
	\end{aligned}
	\end{equation}
Finally, the equation obtained by variation with respect to the two-dimensional metric $g_{ab}$ is 
\begin{equation}\label{eq:eomgab}
	\begin{aligned}
	&\le({\nabla}_a {\nabla}_b - {g}_{ab} {\square}\ri) \Phi^2 + \frac{\Phi}{2} \sum_{i=1}^3 \le( (\partial_a \varphi_i)(\partial_b \varphi_i)  + e^{2\varphi_i} (\partial_a \chi_i )(\partial_b \chi_i)\ri) \\
&+\Bigg[ \frac{1}{\Phi} + \frac{g^2}{2}\Phi\sum_{i=1}^3\le(2\cosh\varphi_i+\chi_i^2e^{\varphi_i}\ri)
- \frac{1}{4\Phi^3} V - \frac{\Phi^2}{4}\sum_{i=1}^3 \le( (\partial \varphi_i)^2  + e^{2\varphi_i} (\partial \chi_i )^2\ri)\Bigg] g_{ab}=0\,.
		\end{aligned}
	\end{equation}
In the next section, we will use these equations of motion to study linear perturbations.

\section{General aspects of the near-\texorpdfstring{AdS$_2$}{AdS2} analysis} \label{sec:generalaspects}
 
 In this section we describe general aspects of the holographic dictionary for near-AdS$_2$ solutions. The aim is to study corners of this dictionary that are sensitive to the surrounding matter content. We will start by constructing the AdS$_2$ backgrounds, which describe the near-horizon geometry of extremal black holes in supergravity. Second, we will discuss the linearized perturbations around the AdS$_2$ solution. This will allow us to identify the JT sector --- which encodes the deviations away from extremality that are characteristic of near-AdS$_2$ --- and the matter degrees of freedom due to the embedding in supergravity.
 And third, we will describe the interactions of the matter fields with the JT sector.
 
 \subsection{\texorpdfstring{AdS$_2$}{AdS2} background: IR fixed point}\label{sec:IR}
 
The AdS$_2$ backgrounds are characterized by having all of the scalars in play equal to a constant: this is the characteristic feature of an attractor mechanism. We will refer to these AdS$_2$ solutions interchangeably as either the attractor solution or the IR fixed point.  Starting with the scalars of our theory in \eqref{eq:steps3}, we will write 
\be\label{eq:attscalar}
\varphi_i = \bar{\varphi}_i~, \quad \chi_i = \bar{\chi}_i~, \quad \Phi = \Phi_0 ~,
\ee
where the right-hand sides are constant values. Inserting this into the dilaton equation \eqref{eq:eomdilaton} gives
	\begin{equation}\label{eq:ricciads2}
		2\Phi_0 R^{(2)} - \frac{2}{\Phi_0^2} + g^2 \sum_i (2\cosh \barp_i + \barc_i^2 e^{\barp_i} ) + \frac{3}{2\Phi_0^4} \bar{V} = 0~, 
	\end{equation}
where $\bar{V}$ indicates that we should evaluate the matrix entries at the fixed point, i.e.\ $\bar V \equiv V({\bf P},{\bf Q})|_{\varphi_i = \bar{\varphi}_i, \chi_i = \bar{\chi}_i}$.  This equation makes it very clear that at the IR fixed point the two-dimensional metric is locally AdS$_2$. To compensate for the odd factors in \eqref{eq:4d2d}, we will set 
\be
 g_{ab} = \Phi_0\, \bar g_{ab} ~,
 \ee
where $\bar g_{ab}$ is a locally AdS$_2$ spacetime with radius $\ell_2$.
Combining \eqref{eq:ricciads2} with the Einstein equations \eqref{eq:eomgab}, we obtain
	\begin{equation}\label{eq:adsphi}
	\begin{aligned}
		\frac{1}{\ell_2^2} + \frac{1}{\Phi_0^2}&=   \frac{1}{2\Phi_0^4}\bar{V} ~,\\
		\frac{1}{\ell_2^2} - \frac{1}{\Phi_0^2}&= g^2 \sum_i (2\cosh \barp_i + \barc_i^2 e^{\barp_i} )~.
	\end{aligned}
	\end{equation}
Using \eqref{eq:attscalar}, the equations for the scalars in \eqref{eq:eomscalars} are
	\begin{equation}\label{att:varphi}
		g^2 (2\sinh{\barp_i} + \barc_i^2 e^{\barp_i}) - \frac{1}{2\Phi_0^4} {\partial_{\barp_i} \bar V} = 0~,
	\end{equation}
and 
\begin{equation}\label{att:chi}
		2g^2 \barc_i  e^{\barp_i}- \frac{1}{2\Phi_0^4} {\partial_{\barc_i} \bar V} = 0~.
	\end{equation}
where the derivatives of the potential are simply
\be
\partial_{\barp_i} \bar V \equiv \frac{\partial V}{\partial \varphi_i}\Big|_{\varphi_i=\barp_i, \chi_i = \bar{\chi}_i}~,
\ee
and similar for ${\partial_{\barc_i} \bar V}$ and multiple derivatives of the potential. 

In App.\,\ref{app:ads2} we write explicitly how the attractor equations \eqref{att:varphi}-\eqref{att:chi} depend on $({\bf P}^I,{\bf Q}_I)$, which illustrates more clearly how $\barp_i$ and $\barc_i$ depend on the charges. In Sec.\,\ref{sec:ungauged} we discuss the solutions of the ungauged case, and in Sec.\,\ref{sec:gauged-cases}  we solve these equations explicitly for special cases in the gauged theory. 

It is important to stress that until this point we have not imposed supersymmetry on the background AdS$_2$ solution: we are only demanding that we have an extremal solution. As we explore explicit cases in gauged ($g^2>0$) and ungauged supergravity ($g=0$), we will describe how imposing BPS conditions on the background affects our subsequent analysis of the near-AdS$_2$ dynamics. 
 
\subsection{Linear analysis: spectrum of operators and JT sector}\label{sec:lineargen}
The first entry in the holographic analysis we will decode  are the linear fluctuations around the AdS$_2$ background, and we will identify the degrees of freedom. Following similar steps as in \cite{Castro:2018ffi,Castro:2021fhc}, we define
 	\begin{equation}\label{def:fluctuations}
	\begin{aligned}
		\Phi &= \Phi_0 + \Y ~, \\
		\varphi_i &= \bar{\varphi}_i + \bph_i ~, \\
		\chi_i &= \bar{\chi}_i + \bch_i~, \\
		g_{ab} &=  \Phi_0\,\bar{g}_{ab} +  h_{ab}~,
	\end{aligned}
	\end{equation}
where $(\Phi_0,\bar{\varphi}_i,\bar{\chi}_i,\bar{g}_{ab})$ define the zeroth-order AdS$_2$ background of Sec.\,\ref{sec:IR}, and $(\Y,\bph_i, \bch_i,h_{ab})$ are the corresponding fluctuations.
 
At the linear level in the fluctuations, using \eqref{def:fluctuations} in the equations of motion gives the following results. Upon using the zeroth-order equations for the background, at leading order the Einstein equations \eqref{eq:eomgab} give 
	\begin{equation}\label{eq:jt1} 
		\le(\bar{\nabla}_a \bar{\nabla}_b - \bar{g}_{ab} \bar{\square}\ri)\Y + \frac{1}{\ell_2^2} \bar{g}_{ab} \Y = 0~.
	\end{equation}	
As expected, this defines $\Y$ to comply with the equation of motion characteristic of JT gravity. The trace of this equation implies that 
\be\label{eq:boxy}
\bar\square\Y= \frac{2}{\ell_2^2}\Y~,
\ee
which identifies $\Y$ as an operator of conformal dimension $\Delta_\Y=2$.
Next, the linearized equations derived from \eqref{eq:eomscalars} are given by
 	\begin{equation} \label{eq:linearphi}
	\begin{aligned}
		0=   \bar{\square} \bph_i  &+ \left( g^2 (2\cosh{\barp_i}+ \barc_i^2 e^{\barp_i} ) - \frac{1}{\ell_2^2} - \frac{1}{\Phi_0^2} \right) \bph_i + \frac{4g^2}{\Phi_0} \left( 2\sinh \barp_i + \barc_i^2 e^{\barp_i} \right) \Y \\
		& \qquad - \frac{1}{2\Phi_0^4}  \sum_{j\neq i} \left( {\partial_{\barp_j}\partial_{\barp_i} \bar V} \bph_j +{\partial_{\barc_j}\partial_{\barp_i} \bar V} \bch_j \right) ~,
	\end{aligned}
	\end{equation}		
and
	\begin{equation} \label{eq:linearchi}
	\begin{aligned}
		0=  e^{2\barp_i} \bar{\square} \bch_i &+ \left( 2g^2 e^{\barp_i} - \frac{1}{2\Phi_0^4} {\partial_{\barc_i}^2 \bar V} \right) \bch_i + \frac{8g^2}{\Phi_0} \barc_i e^{\barp_i} \Y 
		\\ &\qquad 
		- \frac{1}{2\Phi_0^4} \sum_{j\neq i} \left( {\partial_{\barc_j}\partial_{\barc_i} \bar V} \bch_j + {\partial_{\barp_j}\partial_{\barc_i} \bar V} \bph_j \right)~.
	\end{aligned}
	\end{equation}
It is worth noting the different behaviour of the fluctuations for the gauged versus ungauged theory. For $g=0$, the fluctuations of $(h_{ab},\Y)$ are decoupled from the matter sector $(\bph_i,\bch_i)$.  For $g\neq0$ we have to decouple further the fluctuations. We will do so by splitting the solutions of \eqref{eq:linearphi} and \eqref{eq:linearchi} into a homogeneous and inhomogeneous part: 
	\begin{equation}
	\begin{aligned}\label{eq:hominh}
		\bph_i &= \bph_i^{\rm hom} + \frac{a_{\varphi,i}}{\Phi_0}\Y ~, \\
		\bch_i &= e^{-\bar\varphi_i}\bch_i^{\rm hom}  + e^{-\bar\varphi_i}\frac{a_{\chi,i}}{\Phi_0} \Y~.
	\end{aligned}
	\end{equation}
The homogeneous terms in \eqref{eq:hominh} satisfy at linear order
	\begin{equation}\label{eq:linearphihom1}
		\begin{aligned}
		  \bar{\square} \bph_i^{\rm hom}  = &\left(  \frac{1}{\ell_2^2}+ \frac{1}{\Phi_0^2}- g^2 (2\cosh{\barp_i}+ \barc_i^2 e^{\barp_i} ) \right) \bph_i^{\rm hom}\\ &+ \frac{1}{2\Phi_0^4}  \sum_{j\neq i} \left( {\partial_{\barp_j}\partial_{\barp_i} \bar V} \bph_j^{\rm hom} + e^{-\barp_j}{\partial_{\barc_j}\partial_{\barp_i} \bar V} \bch_j^{\rm hom} \right) ~,
		  	\end{aligned}
	\end{equation}
and
	\begin{equation}\label{eq:linearchihom1}
	\begin{aligned}
		  \bar{\square} \bch_i^{\rm hom} = 
		  &\left(  \frac{1}{\ell_2^2} + \frac{1}{\Phi_0^2} -g^2 (2\cosh{\barp_i}+ \barc_i^2 e^{\barp_i} ) \right)\bch_i^{\rm hom} \\&+ \frac{1}{2\Phi_0^4} \sum_{j\neq i} \left( {e^{-\barp_i-\barp_j}\partial_{\barc_j}\partial_{\barc_i} \bar V}\, \bch_j^{\rm hom}+ e^{-\barp_i}{\partial_{\barp_j}\partial_{\barc_i} \bar V} \,\bph_j^{\rm hom} \right) ~.
	\end{aligned}
	\end{equation}
 It will be convenient to introduce some further notation to encode the information of these linear equations. We will write 
 \eqref{eq:linearphihom1}-\eqref{eq:linearchihom1} as
 \be\label{eq:linear-matter}
 \bar{\square} \vec{\bpz}_{\rm h} = \mathfrak{M}^2 \vec{\bpz}_{\rm h}  ~,
 \ee
 where $\vec{\bpz}_{\rm h}\equiv(\bph_i^{\rm hom},\bch_i^{\rm hom})$, and $\mathfrak{M}^2$ is a $6\times6$ mass matrix that can be read off from the above equations. 
  The coefficients $a_{\varphi,i}$ and $a_{\chi,i}$, which parametrize the inhomogeneous solution, are determined such that terms proportional to $\Y$ in \eqref{eq:linearphi} and \eqref{eq:linearchi} are removed from the equations. The condition is
 \be\label{eq:avec}
 \le(\mathfrak{M}^2-\frac{2}{\ell_2^2}\mathds{1}_{6\times6}\ri) \vec{a} = - 4g^2\vec{b}~,\qquad \vec{b}\equiv\begin{pmatrix}
   (2\sinh\barp_i + \barc_i^2 e^{\barp_i})\\
  2 \barc_i 
 \end{pmatrix}~,
 \ee
 with $\vec a= (a_{\varphi,i},a_{\chi,i})$. We will solve for $\vec a$ explicitly for the cases in Sec.\,\ref{sec:gauged-cases}. Notice that for $g=0$, the solution to \eqref{eq:avec} is $\vec a =0$, reflecting that there is no inhomogeneous solution in that case. 
 
Finally, from \eqref{eq:eomdilaton} we obtain the linearized equation for the metric fluctuation, $h_{ab}$, which reads
	\begin{equation}
	\begin{aligned}\label{eq:ricciads3}
		 & - \bar{R}^{ab} h_{ab} + \bar{\nabla}^a\bar{\nabla}^b h_{ab} - \bar{\square}  h^a_{~a} - \left( \frac{8}{\ell_2^2} + \frac{4}{\Phi_0^2} \right) \Y \\
		 & \qquad +\, 2g^2 \Phi_0 \sum_{i=1}^3 \le( \le( 2\sinh\bar{\varphi}_i+(\bar{\chi}_i)^2e^{\bar{\varphi}_i} \ri) \bph_i + 2 e^{\bar{\varphi}_i} \bar{\chi}_i \bch_i \ri)= 0 ~.
	\end{aligned}
	\end{equation}
 This equation reflects how the metric mixes with the JT field $\Y$, which is a standard feature of sphere reductions, and for $g\neq0$, how the matter fields get intertwined as well. To disentangle this equation, and identify the inhomogeneous piece $H^{\rm inh}$, we start by writing 
 \be\label{eq:inhmetric}
h_{ab}=   \hat h_{ab}^{\rm ST}+ \frac{1}{2}\bar g_{ab} \hat h  +\Phi_0\,\bar g_{ab} \, H^{\rm inh}~, 
 \ee
where $\hat h_{ab}^{\rm ST}$ is a symmetric traceless tensor; and $\hat h$, together with $H^{\rm inh}$, describe the trace of the perturbation.     It is simple to check that the inhomogeneous solution  to \eqref{eq:ricciads3} is
\be
H^{\rm inh} = \frac{1}{2} \vec{a}\cdot \vec{\bpz}_{\rm h}~,
\ee
with $\vec{a}$ given by \eqref{eq:avec}, and  
\be\label{eq:traceeomh}
\bar\square\hat h= \frac{2}{\ell_2^2}\hat h~.
\ee
But notice that $\hat h$ is not a new degree of freedom: this equation reflects the residual diffeomorphism that we can still do on the background metric \cite{Michelson:1999kn}. 
Finally, from \eqref{eq:ricciads3}, the term  $\hat h_{ab}^{\rm ST}$ obeys
 	\begin{equation}
	\begin{aligned}\label{eq:ricciads4}
		 & \bar{\nabla}^a\bar{\nabla}^b \hat h_{ab}^{\rm ST} - \left( \frac{8}{\ell_2^2} + \frac{4}{\Phi_0^2} - 2g^2  \vec{b}\cdot \vec{a}\right) \Y 
		 = 0 ~,
	\end{aligned}
	\end{equation}
 which couples the metric to $\Y$ and we will discuss the solutions when addressing the JT sector.  
 
To summarize, the independent fluctuations around the AdS$_2$ backgrounds correspond to the JT field $\Y$, and the matter sector containing $(\bph^{\rm hom}_i,\bch^{\rm hom}_i)$. For each matter degree of freedom we will have a corresponding dual operator; and the six eigenvalues of the mass matrix in \eqref{eq:linear-matter} will be related to the conformal dimensions of these operators in the standard way,
 \be
\Delta_{i} = \frac{1}{2}\le(1+\sqrt{1+4 m^2_{i} \ell_2^2}\ri)~,
 \ee
with $m^2_{i}$ an eigenvalue of $\mathfrak{M}^2$. This, plus $\Delta_\Y=2$, corresponds to the spectrum of operators in our system.

\paragraph{JT sector.} This is the portion of the fluctuations that is controlled by $\Y$, and the homogeneous solutions are trivial. More explicitly, in \eqref{def:fluctuations} we set
 	\begin{equation}
	\begin{aligned}
		(\bph_i, e^{\bar\varphi_i} \bch_i)&=     \frac{\Y}{\Phi_0}\vec{a}~,\\
		h_{ab}&=   \hat h_{ab}^{\rm ST}~,
	\end{aligned}
	\end{equation}
with $\vec{a}$ the solution to \eqref{eq:avec}, $\hat h^{\rm ST}_{ab}$  satisfies \eqref{eq:ricciads4}, and $\Y$ is governed by \eqref{eq:jt1}. The dynamical aspects of $\Y$,  defined by its distinctive equation of motion \eqref{eq:jt1}, are well described by JT gravity. Provided some assumptions on the operator content of the theory, this sector controls the deviations away from extremality, where it can be seen that a non-trivial profile of $\Y$ accounts for the response of the black hole as the temperature is increased \cite{Almheiri:2014cka,Maldacena:2016hyu}. This has been well reported for spherically symmetric cases in four dimensions in \cite{Almheiri:2016fws,Nayak:2018qej} when the theory was only Einstein-Maxwell theory. 

To complete the discussion of the JT sector, we will construct the solutions to \eqref{eq:ricciads4}. We will write the traceless part as
 \be\label{eq:hst}
\hat h_{ab}^{\rm ST}= \bar\nabla_a \bar\nabla_b U(x) -\frac{1}{2}g_{ab}\bar\square U(x)~,
\ee 
with $U(x)$ a scalar function. Then \eqref{eq:ricciads4} reduces to 
\be
(\bar\square-\frac{2}{\ell_2}) U(x) = \left( \frac{4}{\ell_2^2} + \frac{2}{\Phi_0^2} - g^2  \vec{b}\cdot \vec{a}\right) \Y~. 
\ee
Although we are describing the solution to $\hat h_{ab}^{\rm ST}$ in the context of the linearized equations, it is important to note that this backreaction of the geometry is a higher-order effect in powers of $\Y$. The reason is simply that \eqref{eq:ricciads4} is obtained by varying the action with respect to $\Y$, and this requires terms that are schematically of the form $h\Y + \Y^2$. We will only focus on linear order effects in $\Y$ and hence \eqref{eq:hst} will not play a role in Sec.\,\ref{sec:interact}.   

To close the analysis of this subsection, it is instructive to compare with prior results. In particular, \cite{Michelson:1999kn,Larsen:2014bqa,Larsen:2018cts} contains a detailed study of the linearized spectrum of AdS$_2\times S^2$ in the ungauged theory, which includes all possible harmonics on $S^2$. Our analysis corresponds to the lowest $l=0$ sector in their notation.  One important difference is that  \cite{Michelson:1999kn,Larsen:2014bqa} makes a choice of Lorentz, or de Donder, gauge in four dimensions; here we have not fixed a gauge. This choice of gauge forces $\Y=0$ and the JT dynamics appear as boundary modes. Otherwise our analysis is in agreement with theirs. 
 
\subsection{Interactions in near-\texorpdfstring{AdS$_2$}{AdS2}} \label{sec:interact}

Finally, we will focus on the interactions between $\Y$ and the  matter fields in our model.  We will be treating $\Y$ as a background field --- with its non-trivial profile driving the system away from the ideal AdS$_2$ background by explicitly breaking conformal invariance --- and we will quantify its imprint on the matter sector. In such a  scenario, cubic interactions that involve one power of $\Y$ capture the leading correction to the two-point functions which we aim to evaluate.   
 
In this subsection we will only describe the general aspects of these interactions, and how they enter in the two-point function of the matter sector; in the subsequent sections we will evaluate these corrections explicitly for specific cases.  The general strategy presented here follows the discussion in \cite{Castro:2021fhc}, which we summarize and apply to our situation. 

To quantify these interactions, and their impact on two-point functions of the matter fields, we start by writing the bosonic supergravity fields as 
	\begin{equation}\label{eq:fluctuations}
	\begin{aligned}
	\Phi&= \Phi_0 + \Y~,\\
		(\varphi_i, e^{\bar\varphi_i} \chi_i)&=(\bar\varphi_i, e^{\bar\varphi_i} \bar\chi_i)+  \vec{a}  {\Y\over \Phi_0} + (\bph_i, e^{\bar\varphi_i} \bch_i)~,\\
		g_{ab}&=\Phi_0\, \bar g_{ab}+  {\Phi_0\over 2} \,\bar g_{ab}\, \vec{a}\cdot \vec{\bpz} ~.
	\end{aligned}
	\end{equation}
Here $(\bar g_{ab},\Phi_0,\bar\varphi_i,\bar\chi_i)$ specifies the AdS$_2$ background, which complies with the equations of motion in Sec.\,\ref{sec:IR}; the vector $\vec a$ is defined in \eqref{eq:avec}. We also have $\vec{\bpz}\equiv (\bph_i, \bch_i)$ which describes the matter degrees of freedom beyond the linearized level discussed in Sec.\,\ref{sec:lineargen}.

Starting from \eqref{eq:steps3}-\eqref{eq:defV}, 
we will build the effective action that captures the dynamics of  $\vec{\bpz}$, including interactions with the JT sector to leading order in $\Y$. This effective Euclidean action will be of the form
    \be
    \begin{aligned}
        I_\text{eff}&= {\Phi_0^2\over 4G_4}\int\! d^2x\sqrt{\bar{g}}\le( {\cal L}_{\rm free} + {\cal L}_{\rm int}\ri)~,\label{act:twop}
    \end{aligned}
    \ee
where the free portion describes the quadratic action for  $\vec{\bpz}$ 
 \be \label{eq:kinActiongauged}
  \begin{aligned}
 {\cal L}_{\rm free}=\frac{1}{2} \partial_a \vec{\bpz} \cdot \partial^a \vec{\bpz} +\frac{1}{2} \vec{\bpz}^T\, \mathfrak{M}^2\, \vec{\bpz} ~,
\end{aligned}
 \ee
with the mass matrix as given by \eqref{eq:linear-matter}. In obtaining $ {\cal L}_{\rm free}$ it was important to introduce terms proportional to $\vec a$ in \eqref{eq:fluctuations}, since otherwise  the matter degrees of freedom in $\vec{\bpz}$ would not decouple from the gravitational degrees of freedom  $(\Y, h_{ab})$. As expected, the free terms capture the homogeneous solutions of the linearized spectrum in Sec.\,\ref{sec:lineargen}.  
The leading cubic interaction terms are\footnote{About notation: $\bpz_i$ are the components of $\vec \bpz$, and explicitly we have $(\bpz_1,\bpz_2,\bpz_3,\bpz_4,\bpz_5,\bpz_6)=(\bph_1,\bph_2,\bph_3,\bch_1,\bch_2,\bch_3)$.} 
  \be
    \begin{aligned}
       {\cal L}_{\rm int}= & \frac{1}{2}\partial_a \Y (\partial^a\vec{\bpz}\cdot \vec{a}) \vec{\bpz}\cdot \vec{a}+ \Y \partial_a \vec{\bpz} \cdot \partial^a \vec{\bpz}\\
&+ \frac{1}{2}\sum^3_{i=1} \le(a_{\chi,i} \bph_i \partial_a\Y\partial^a\bch_i  +2 a_{\varphi,i} \Y\partial^a\bch_i\partial_a\bch_i\ri)\\ &+ \sum_{i,j=1}^6\lambda_{{\bpz}_i{\bpz}_j\mathcal{Y}}\, {\bpz}_i\,{\bpz}_j\,{\mathcal{Y}}~.
    \end{aligned}
    \ee
The expression of $\lambda_{{\bpz}_i{\bpz}_j\mathcal{Y}}$, in terms of the background AdS$_2$ solution and $\vec a$, is rather lengthy for \eqref{eq:steps3}, and hence  we are not writing it explicitly. But it is straightforward to evaluate.  

It will also be useful to bring \eqref{act:twop} to a diagonal basis, where it is simple to read off the eigenvalues of $\mathfrak{M}^2$. For this reason we define $\vec{\mathfrak{Z}}$, which contains the orthogonal eigenstates of $\mathfrak{M}^2$. In this basis \eqref{act:twop} becomes 
 \be \label{eq:kinActiongauged-diag}
  \begin{aligned}
 I_{\rm eff}=&\int\! d^2x\sqrt{\bar{g}}\bigg( \frac{1}{2} \partial_a \vec{\mathfrak{Z}} \cdot \partial^a \vec{\mathfrak{Z}} +\frac{1}{2} \vec{\mathfrak{Z}} M^2 \vec{\mathfrak{Z}} \\&+ \sum_{i,j}\le(\lambda_{{\mathfrak{Z}_i}{\mathfrak{Z}_j}\mathcal{Y}} \,{\mathfrak{Z}}_i\,{\mathfrak{Z}}_j\,\mathcal{Y}
  +{\lambda}_{{\mathfrak{Z}_i}(\p{\mathfrak{Z}_j})(\p\mathcal{Y})}{\mathfrak{Z}_i}\p_a{\mathfrak{Z}_j}\p^a\mathcal{Y}+
  {\lambda}_{\mathcal{Y}(\p{\mathfrak{Z}_i})(\p{\mathfrak{Z}_j})}
  \mathcal{Y}\p_a{\mathfrak{Z}_i}\p^a{\mathfrak{Z}_j}\ri)\bigg)~.
\end{aligned}
 \ee
Here $M^2={\rm diag}(m_1^2,...,m_6^2)$ contains in its diagonal the eigenvalues of $\mathfrak{M}^2$, and the cubic couplings appearing above  will follow. Also note that $\vec{\mathfrak{Z}}$ has been appropriately normalized to remove the overall factor of ${\Phi_0^2\over 4G_4}$ in \eqref{act:twop}. 

Next let us focus on the evaluation of the two-point function of one of the eigenstates in \eqref{eq:kinActiongauged-diag}, which for sake of simplicity (and abusing notation) we will call $\mathfrak{Z}(x)$, and whose mass eigenvalue within $M^2$ is $m^2_{{\mathfrak{Z}}}$. 
To start, in the vacuum (pure AdS in the bulk), we can use coordinates such that
\be\label{eq:poincare}
ds_{\text{AdS}_2}^2=\frac{\ell_2^2}{z^2}(\dd t^2+ \dd z^2)\,.
\ee
 In these coordinates, the near-boundary expansion of ${\mathfrak{Z}}$ is 
\be\label{eq:assXhat}
{\mathfrak{Z}}(t,z)=z^{1-\Delta}\tilde{\mathfrak{Z}}(t)+\cdots,\qquad\text{as}\,\,z\to0\,,
\ee
and $\tilde{\mathfrak{Z}}(t)$ is interpreted as the source of its dual operator $\mathcal{O}_{{\mathfrak{Z}}}(t)$ with dimension
\be
\Delta \equiv \Delta_{\mathfrak{Z}} = \frac{1}{2}\le(1+\sqrt{1+4 m^2_{{\mathfrak{Z}}} \ell_2^2}\ri)\,.
\ee
Next, we can perform a diffeomorphism in the bulk to go to a thermal state, with inverse temperature $\beta$. Near the boundary, the appropriate thermal transformation is
\be\label{eq:therm}
t(u)=\tan\left(\frac{\pi}{\beta}u\right)\,,\qquad u\sim u+\beta\,,
\ee
with $u$ being the boundary time. As we do this transformation, we also want to keep track of the UV cuttoff and make sure the proper length of the boundary is fixed, i.e.  
\be
g|_\text{bndy}=\frac{\ell_2^2}{\epsilon^2}~.
\ee
The above condition implies that
\be
z(u)=\epsilon\sqrt{(t')^2+(z')^2}=\epsilon t(u)' +\mathcal{O}(\epsilon^3)\,,
\ee
fixing the asymptotic part of the radial part of the diffeomorphism. 
Then, under this transformation, the asymptotic form of the field ${\mathfrak{Z}}$ becomes
\be
{\mathfrak{Z}}(t,z)=\epsilon^{1-\Delta}[t'(u)]^{1-\Delta}\tilde{\mathfrak{Z}}(t(u))+\cdots,\qquad\text{as}\,\,\epsilon\to0\,,
\ee
and now $[t'(u)]^{1-\Delta}\tilde{\mathfrak{Z}}(t(u))\equiv\bar{\mathfrak{Z}}(u)$ is interpreted as a source. 

Turning next to the behavior of $\Y$, this term will be treated as a background field whose source is non-trivial. More explicitly, we will take the on-shell value\footnote{The addition of $\Phi_0^{-1}$ in \eqref{eq:expPHI} is to make the field $\Y$ dimensionless.}
\be\label{eq:expPHI}
\Phi_0^{-1}\Y(t,z)=z^{-1} \tilde{\mathcal{Y}}(t) + \cdots 
\ee
and set the source on the thermal state fixed, i.e.\ $\bar{\mathcal{Y}}(u)=a$ with $a$ a constant.  This defines a near-AdS$_2$ background. 
Then, evaluating the two-point functions of the operator dual to $\mathfrak{Z}$  in the presence of  a background value for $\Y$ can be done by treating $\Y$ as an operator with $\Delta=-1$, and integrating over its boundary time \cite{Maldacena:2016upp}. 
More explicitly, the thermal two-point function we are after is, to leading order in $\Y$,
\be
\langle\mathcal{O}_\mathfrak{Z}(u_1)\mathcal{O}_\mathfrak{Z}(u_2)\rangle_\beta=\langle\mathcal{O}_\mathfrak{Z}(u_1)\mathcal{O}_\mathfrak{Z}(u_2)\rangle^{\text{free}}_\beta + \langle\mathcal{O}_\mathfrak{Z}(u_1)\mathcal{O}_\mathfrak{Z}(u_2)\rangle^{\text{int}}_\beta~,
\ee
where the correction due to $\Y$ is
\begin{align}
\langle\mathcal{O}_\mathfrak{Z}(u_1)\mathcal{O}_\mathfrak{Z}(u_2)\rangle^{\text{int}}_{\beta}&=a\int_0^\beta \dd  u_3\,\langle\mathcal{O}_\mathfrak{Z}(u_1)\mathcal{O}_\mathfrak{Z}(u_2)\mathcal{O}_{-1}(u_3)\rangle_{\beta}~.
\end{align}

We can now easily evaluate the appropriate two-point function. As standard in AdS/CFT, the boundary effective action on the vacuum \eqref{eq:poincare} is:
\be
I_{\text{eff}}=I_{\text{free}}+I_{\text{interactions}}\,.
\ee
The free part yields
\be\label{eq:defD}
I_\text{free}=- \frac{D}{2}\int \dd t_1 \dd t_2\frac{\tilde{\mathfrak{Z}}(t_1) \tilde{\mathfrak{Z}}(t_2)}{|t_1-t_2|^{2\Delta_{\mathfrak{Z}}}}\,,\qquad D= \frac{(2\Delta - 1)\Gamma[\Delta]}{\sqrt{\pi}\Gamma[\Delta-\frac{1}{2}]}\,.
\ee
The thermal two-point function is 
\be\label{eq:twopfree}
\langle\mathcal{O}_ \mathfrak{Z}(u_1)\mathcal{O}_ \mathfrak{Z}(u_2)\rangle^{\text{free}}_\beta= D\left[\frac{t'(u_1)t'(u_2)}{|t(u_1)-t(u_2)|^{2}}\right]^\Delta= D\left[\frac{\pi}{\beta\sin(\frac{\pi u_{12}}{\beta})}\right]^{2\Delta}\!\!,
\ee
where $u_{12}\equiv u_1-u_2$.
For the cubic interactions we will have 
\be\label{intAct}
\begin{aligned}
I_{\text{interactions}}&=\frac{\tilde{D}}{2} \int \frac{\dd t_1 \dd t_2 \dd t_3\,\tilde{\mathfrak{Z}}(t_1)\tilde{\mathfrak{Z}}(t_2)\tilde{\mathcal{Y}}(t_3)}{|t_{12}|^{2\Delta+1}|t_{23}|^{-1}|t_{31}|^{-1}}\\
&=\frac{\tilde{D}}{2}\int \dd u_1 \dd u_2 \dd u_3\frac{t'(u_1)^\Delta t'(u_2)^\Delta t'(u_3)^{-1}\bar{\mathfrak{Z}}(u_1)\bar{\mathfrak{Z}}(u_2)\bar{\mathcal{Y}}(u_3)}{|t(u_1)-t(u_2)|^{2\Delta+1}|t(u_1)-t(u_3)|^{-1}|t(u_2)-t(u_3)|^{-1}}
\,,
\end{aligned}
\ee
where 
\be\label{defTilD}
\tilde{D}\equiv\lambda_{\mathcal{Y}{\mathfrak{Z}}{\mathfrak{Z}}}K_{\mathcal{Y}{\mathfrak{Z}}{\mathfrak{Z}}}+{\lambda}_{ {\mathfrak{Z}}(\p {\mathfrak{Z}})(\p\mathcal{Y})}{K}_{ {\mathfrak{Z}}(\p {\mathfrak{Z}})(\p\mathcal{Y})} +{\lambda}_{\mathcal{Y}(\p{\mathfrak{Z}})(\p{\mathfrak{Z}})}{K}_{\mathcal{Y}(\p{\mathfrak{Z}})(\p{\mathfrak{Z}})}~,
\ee
and the coefficients appearing here are \cite{Freedman:1998tz}
\be\label{Kcoeff}
\begin{aligned}
K_{\mathcal{Y} {\mathfrak{Z}} {\mathfrak{Z}}}&=-\frac{\Gamma[-\frac{1}{2}]^2\Gamma[\Delta+\frac{1}{2}]\Gamma[\Delta-1]}{2\pi\Gamma[\Delta-\frac{1}{2}]^2\Gamma[-\frac{3}{2}]}=
-\frac{3 (\Delta -\frac{1}{2}) \Gamma [\Delta -1]}{2 \sqrt{\pi } \Gamma [\Delta -\frac{1}{2}]}\,,\\
{K}_{ {\mathfrak{Z}}(\p {\mathfrak{Z}})(\p\mathcal{Y})}&=\left[-\Delta+\frac{1}{2}(2-2\Delta)(-1)\right]{K_{\mathcal{Y} {\mathfrak{Z}} {\mathfrak{Z}}}}{\ell_2^2}=-\frac{K_{\mathcal{Y} {\mathfrak{Z}} {\mathfrak{Z}}}}{\ell_2^2} ~,\\
\tilde{K}_{\mathcal{Y}(\p {\mathfrak{Z}})(\p {\mathfrak{Z}})}&=\left[\Delta^2+\frac{1}{2}(2-2\Delta)(2\Delta+1)\right] \frac{K_{\mathcal{Y} {\mathfrak{Z}} {\mathfrak{Z}}}}{\ell_2^2}= - (\Delta^2 - \Delta - 1) \frac{K_{\mathcal{Y} {\mathfrak{Z}} {\mathfrak{Z}}}}{\ell_2^2}~.
\end{aligned}
\ee
Then, the three-point function we need is 
\begin{equation} \label{eq:threep}
\begin{aligned}
\langle\mathcal{O}_ \mathfrak{Z}(u_1)\mathcal{O}_ \mathfrak{Z}(u_2)\mathcal{O}_{-1}(u_3)\rangle_{\beta} 
&=\tilde{D}\frac{ \beta}{\pi}\left[\frac{\pi}{\beta\sin(\frac{\pi u_{12}}{\beta})}\right]^{2\Delta}\frac{|\sin(\frac{\pi u_{13}}{\beta})|\,|\sin(\frac{\pi u_{23}}{\beta})|}{|\sin(\frac{\pi u_{12}}{\beta})|}\,,
\end{aligned}
\end{equation}
which contributes to the correlator as
\begin{align}
\langle\mathcal{O}_ \mathfrak{Z}(u_1)\mathcal{O}_ \mathfrak{Z}(u_2)\rangle^{\text{int}}_{\beta}&=a\int_0^\beta \dd  u_3\,\langle\mathcal{O}_ \mathfrak{Z}(u_1)\mathcal{O}_ \mathfrak{Z}(u_2)\mathcal{O}_{-1}(u_3)\rangle_{\beta}\nonumber\\
&=\tilde{D}\frac{a \beta^2}{2\pi^2}\left[\frac{\pi}{\beta\sin(\frac{\pi u_{12}}{\beta})}\right]^{2\Delta}\left(2+\pi\frac{1-2u_{12}/\beta}{\tan(\frac{\pi u_{12}}{\beta})}\right)\,.\label{two:corr}
\end{align}
See \cite{Castro:2021fhc} for details about evaluating this integral. Adding the free and interaction pieces of the correlator, we find
\be\label{two:total}
\langle\mathcal{O}_ \mathfrak{Z}(u_1)\mathcal{O}_ \mathfrak{Z}(u_2)\rangle_{\beta}=\left[\frac{\pi}{\beta\sin(\frac{\pi u_{12}}{\beta})}\right]^{2\Delta}\left[D+\tilde{D}\frac{ a\beta^2}{2\pi^2}\left(2+\pi\frac{1-2u_{12}/\beta}{\tan(\frac{\pi u_{12}}{\beta})}\right)\right].
\ee
The constants $D$ and $\tilde{D}$ are given in \eqref{eq:defD} and \eqref{defTilD}, respectively.

As we investigate examples in the subsequent sections, one of our main aims will be to report on the value of $\tilde{D}$ compared to $D$. For this purpose it is convenient to define their ratio as
    \begin{equation} \label{eq:Dhat}
        \hat D \equiv \frac{\tilde D}{D} = \frac{\lambda_{\rm eff}K_{\Y \mathfrak{Z}\mathfrak{Z}}}{D} = - \frac{3}{4(\Delta - 1)} \lambda_{\rm eff}~,
    \end{equation}
where $\lambda_{\rm eff} = \lambda_{\rm eff} (\Delta) $ is the effective cubic coupling constant, which depends on the conformal dimension of $\mathfrak{Z}$ via \eqref{Kcoeff}. We will see that the value of $\hat D$ behaves differently depending on the background and surrounding: BPS versus non-BPS backgrounds, gauged versus ungauged theories. Understanding what are possible behaviours is valuable information to understand how specific properties of the dual system accommodate for the properties of the gravitational background.

\section{Ungauged supergravity}\label{sec:ungauged}

It is instructive to specialize to the case $g=0$, and inspect more closely the responses and fluctuations of extremal black holes. The main goal here is two-fold: first, to recover the aspects of the nAttractor mechanism, and second, to contrast the response of near-AdS$_2$ of BPS versus non-BPS black holes. We will start this section by describing the BPS and non-BPS branches; subsequently, we will describe the IR fixed point and the linear analysis, classifying the solutions according to them being BPS or non-BPS. Finally, we study the interactions between the scalars and axions and the dilaton field $\Y$, highlighting the differences between both branches. 

There are multiple references to characterize the ${\cal N}=2$ ungauged supergravity theories at hand, and several different conventions used in those references which we will not summarize here. By setting $g=0$ for the theory presented in Sec.\,\ref{sec:gaugedsugra} one obtains STU supergravity described according to the conventions in \cite{Chow:2014cca}, and those are the conventions used here. 

\subsection{BPS versus non-BPS branch}
For ungauged black holes it is interesting to distinguish, among extremal black holes, which ones preserve supersymmetry (BPS branch) and which do not (non-BPS branch). For regular and static black holes this distinction is elegantly dictated by the Cayley hyperdeterminant. This quantity is the quartic invariant of STU supergravity, and it is defined as \cite{Kallosh:1996uy,Duff:2006uz}
\be\label{eq:hypercayley}
\hat{\mathbf{\Delta}}\equiv \frac{1}{16} \le( 4 Q_1 Q_2Q_3Q_4 +4 P^1 P^2 P^3P^4 + 2\sum_{J<K}Q_JQ_K P^JP^K - \sum_{I} (Q_IP^I)^2\ri)~.
\ee
One of the prominent roles played by this hyperdeterminant is that it controls the area of the extremal black hole and hence the Bekenstein-Hawking entropy of it: 
\be
S_{\rm BH, ext} ={\pi \over G_4} \Phi_0^2 = {2\pi\over G_4} \sqrt{|\hat{\mathbf{\Delta}}|}~.
\ee
 Another utility of this invariant is that we can identify two branches of solutions: an extremal black hole in STU supergravity can then be labelled as
\be
\begin{aligned}
{\rm \bf non-BPS:} \qquad \hat{\mathbf{\Delta}} <0~,\\
{\rm \bf BPS:}\qquad \hat{\mathbf{\Delta}}>0~.
\end{aligned}
\ee
The case $\hat{\mathbf{\Delta}}=0$ is singular for black holes in  a two-derivative supergravity theory since the horizon area vanishes; these configurations are usually referred to as a small black hole. 

\subsection{\texorpdfstring{AdS$_2$}{AdS2} background: IR fixed point}\label{sec:ads2g0}
When $g=0$ the system becomes much more manageable: it is simpler to quantify the background and the fluctuations around it. For the AdS$_2$ background the main simplification comes from the second equation in \eqref{eq:adsphi}, which gives
\be
{\ell_2^2} = {\Phi_0^2}~,
\ee
implying that the AdS$_2$ and $S^2$ radius are equal to each other.  Then, the conditions on the constant scalars supporting the AdS$_2$ gives 
\be\label{eq:atteqnsg0}
{\partial_{\barc_i} \bar V} = {\partial_{\barp_i} \bar V}=0~,
\ee
which are the renowned conditions from the attractor mechanism of extremal black holes; see, for example, \cite{Goldstein:2005hq,Kallosh:2006bt} for a general analysis. Explicitly, they are given by \eqref{eq:atteqsscalars} and \eqref{eq:atteqsaxions} with $g = 0$ for the ungauged theory considered here.

For what follows it will be useful to rewrite the expression for the Cayley hyperdeterminant in terms of the charges dressed by the moduli: $({\cal P}^I, {\cal Q}_I )$ defined in \eqref{eq:calPs}-\eqref{eq:calQsB}. The expression \eqref{eq:hypercayley} becomes 
    \begin{equation}
        \hat{\mathbf{\Delta}} = \frac{e^{-2(\barp_1 + \barp_2 + \barp_3)}}{16}  \Big( 4 {\cal Q}_1 {\cal Q}_2{\cal Q}_3{\cal Q}_4 +4 {\cal P}^1 {\cal P}^2 {\cal P}^3{\cal P}^4 +   2\sum_{J<K}{\cal Q}_J{\cal Q}_K {\cal P}^J{\cal P}^K - \sum_{I} ({\cal Q}_I{\cal P}^I)^2 \Big)~.
    \end{equation}
    It is also simpler to discuss the solutions to the attractor equations \eqref{eq:atteqnsg0} in terms of $({\cal P}^I, {\cal Q}_I )$. 
In this notation, the BPS branch ($\hat{\mathbf{\Delta}}>0$) of solutions is very simple and dictated by linear conditions as follows.  First, a solution to \eqref{eq:atteqnsg0} can be obtained by demanding\footnote{Despite appearances, \eqref{eq:BPSequalPQs} does not imply that the physical charges $(Q_I,P^I)$ are set equal: on the BPS branch, these conditions allow for 4 electric and 4 magnetic independent charges. For solutions on the non-BPS branch that comply with \eqref{eq:BPSequalPQs}, there will be at least one constraint among $(Q_I,P^I)$, and hence this type of solution is not the most general non-BPS configuration.}
    \begin{equation} \label{eq:BPSequalPQs}
       | {\cal P}^1 | = |{ \cal P }^2 | = | {\cal P }^3 | = | {\cal P}^4 | ~, \quad \text{and} \quad |{ \cal Q}_1 | = | {\cal Q}_2 | = | {\cal Q}_3 | = |{ \cal Q}_4 |~.
    \end{equation}
Second, to make this a BPS solution, one has to select an even number of minus signs, and the signs of the ${\cal P}^I$ and ${\cal Q}_I$ are matched. For example, one possible BPS solution is of the form
    \begin{equation}\label{eq:BPSequalPQs1}
        {\cal P}^1 = { \cal P}^2 = - {\cal P}^3 = - { \cal P}^4~, \quad { \cal Q}_1 = { \cal Q}_2 = - {\cal Q}_3 = - {\cal Q}_4~.
    \end{equation}
Without loss of generality,    we will use this choice of signs as a representative of the BPS branch in what follows.  
    
In the non-BPS branch ($\hat{\mathbf{\Delta}}<0$), solutions of the form \eqref{eq:BPSequalPQs} also exist, but with an odd number of relative minus signs for both ${\cal Q}_I$ and ${\cal P}^I$; contrary to the BPS branch, the signs do not have to be matched. One example of this class of solutions is
\be\label{eq:nonBPSequalPQs1}
{\cal P}^1 = -{\cal P}^2 = - {\cal P}^3 = -{\cal P}^4 ~,\quad {\cal Q}_1 =  {\cal Q}_2 = - {\cal Q}_3 =  {\cal Q}_4~.
\ee
However, there exist non-BPS solutions to \eqref{eq:atteqnsg0} that do not comply (or only partially comply)  with the conditions \eqref{eq:BPSequalPQs}. In App.\,\ref{app:scalings}, we will comment on this other class of non-BPS attractor solutions, and how both BPS and non-BPS configurations can be obtained as the extremal limit of the black hole solutions discussed in \cite{Chow:2014cca}. 

Lastly, the AdS$_2$ radius \eqref{eq:adsphi}, or \eqref{eq:L2equation}, in terms of $({\cal P}^I, {\cal Q}_I )$ is given by
\be
{\ell_2^2} =  \frac{1}{4}e^{-\barp_1 - \barp_2 - \barp_3} \Big( {\cal Q}_1^2 + {\cal Q}_2^2 + {\cal Q}_3^2 + {\cal Q}_4^2 + ({\cal P}^4)^2 + ({\cal P}^3)^2 + ({\cal P}^2)^2 +  ({\cal P}^1)^2 \Big) ~.
\ee
And also note that $\ell_2^2=\sqrt{|\hat{\mathbf{\Delta}}|}$. For both BPS and non-BPS solutions of the type \eqref{eq:BPSequalPQs}, we can simplify this expression further, which reads
    \begin{equation}
        \ell_2^2 = e^{-\barp_1 -\barp_2 -\barp_3} \lp ({\cal P}^1)^2 + ({\cal Q}_1)^2 \rp~.
    \end{equation}


\subsection{Linear analysis} \label{sec:linearUngauged}

In this section we revisit the linear analysis done in Sec.\,\ref{sec:lineargen}. As noted there, several simplifications occur on the spectrum of fluctuations when $g=0$; the most prominent one  being that we do not have inhomogeneous terms that mix the JT field with the matter content. In the following we will gather those expressions and solve them for the BPS and non-BPS backgrounds discussed in Sec.\,\ref{sec:ads2g0}. 

We will cast the perturbations as
 	\begin{equation}\label{def:fluctuationsg0}
	\begin{aligned}
		\Phi &= \Phi_0 +  \Y ~, \\
		\varphi_i &= \bar{\varphi}_i + \bph_i^{\rm hom} ~, \\
		\chi_i &= \bar{\chi}_i + e^{-\bar{\varphi}_i}\bch_i^{\rm hom}~, \\
		g_{ab} &=  \Phi_0\,\bar{g}_{ab} +  \hat h_{ab}~, 
	\end{aligned}
	\end{equation}
reflecting that we do not have inhomogeneous terms present in \eqref{eq:hominh} and \eqref{eq:inhmetric}. In the following we will just replace $\bph_i^{\rm hom}\to \bph_i$ and $\bch_i^{\rm hom}\to \bch_i$, since the label is redundant here. 
As usual $\Y$ is the JT field, obeying \eqref{eq:jt1}.   For the metric fluctuations in \eqref{eq:ricciads4} we have
	\begin{equation}
	\begin{aligned}\label{eq:ricciads5}
		 & \bar{\nabla}^a\bar{\nabla}^b \hat h_{ab} - \bar{\square} \hat h^a_{~a} +  {2\over \ell_2^2} \hat h^a_{~a} -\frac{12}{\ell_2^2}  \Y 
		 = 0 ~,
	\end{aligned}
	\end{equation}
and the linear equations for the scalar fields \eqref{eq:linearphi}-\eqref{eq:linearchi} simplify to
 	\begin{equation}
	\begin{aligned}\label{eq:linearg01}
		0&= \left( \bar{\square}  - \frac{2}{\ell_2^2} \right) \bph_i  - \frac{1}{2\ell_2^4}  \sum_{j\neq i} \left( {\partial_{\barp_j}\partial_{\barp_i} \bar V} \bph_j +{\partial_{\barc_j}\partial_{\barp_i} \bar V} \bch_j \right)~,\\
		0&= \left( \bar{\square}  -  \frac{2}{\ell_2^2} \right) \bch_i - \frac{e^{-2\barp_i}}{2\ell_2^4} \sum_{j\neq i} \left( {\partial_{\barc_j}\partial_{\barc_i} \bar V} \bch_j + {\partial_{\barp_j}\partial_{\barc_i} \bar V} \bph_j \right)~.
	\end{aligned}
	\end{equation}

Our main aim is to determine the mass eigenvalues and eigenvectors of \eqref{eq:linearg01}; with this we will quantify the dual operators that should be part of the holographic description of the near-extremal black hole.  Focusing first on the attractor solutions characterized by \eqref{eq:BPSequalPQs}, then we have 
    \begin{equation}
		\bar{\square} \vec{\psi} - \mathfrak{M}^2 \vec{\psi} = 0~,
    \end{equation}
where $\vec{\psi} = (\bph_i, \bch_i)$, and the mass matrix $\mathfrak{M}^2$ is given by 
	\begin{equation}
		\mathfrak{M}^2 = \frac{2}{\ell_2^2} \begin{pmatrix}
			1 & 0 & 0 & 0 & m_{15} & m_{16} \\
			0 & 1 & 0 & m_{24} & 0 & m_{26} \\
			0 & 0 & 1 & m_{34} & m_{35} & 0 \\
			0 & m_{24} & m_{34} & 1 & m_{45} & m_{46} \\
			m_{15} & 0 & m_{35} & m_{45} & 1 & m_{56} \\
			m_{16} & m_{26} & 0 & m_{46} & m_{56} & 1 
		\end{pmatrix}~,
	\end{equation}
with
	\begin{equation} \label{eq:nondiagonalUngauged}
	\begin{split}
		m_{15} &= \alpha ({\cal P}^4 {\cal Q}_2 - {\cal P}^1 {\cal Q}_3)~, \\
		m_{16} &= \alpha ({\cal P}^4 {\cal Q}_3 - {\cal P}^1 {\cal Q}_2)~, \\
		m_{24} &= \alpha ({\cal P}^4 {\cal Q}_1 - {\cal P}^2 {\cal Q}_3)~, \\
		m_{26} &= \alpha ({\cal P}^4 {\cal Q}_3 - {\cal P}^2 {\cal Q}_1) ~,\\
		m_{34} &= \alpha ({\cal P}^4 {\cal Q}_1 - {\cal P}^3 {\cal Q}_2) ~,
	\end{split}
	~\qquad  \qquad
	\begin{split}
		m_{35} &= \alpha ({\cal P}^4 {\cal Q}_2 - {\cal P}^3 {\cal Q}_1)~, \\
		m_{45} &= \frac{\alpha}{2} ({\cal P}^4 {\cal P}^3 - {\cal Q}_1 {\cal Q}_2 - {\cal P}^1 {\cal P}^2 + {\cal Q}_3{\cal Q}_4)~, \\
		m_{46} &= \frac{\alpha}{2} ({\cal P}^4 {\cal P}^2 - {\cal Q}_1 {\cal Q}_3 - {\cal P}^1 {\cal P}^3 + {\cal Q}_2{\cal Q}_4)~, \\
		m_{56} &= \frac{\alpha}{2} ({\cal P}^4 {\cal P}^1 - {\cal Q}_2 {\cal Q}_3 - {\cal P}^2 {\cal P}^3 + {\cal Q}_1{\cal Q}_4) ~,
	\end{split}
	\end{equation}
where 
	\begin{equation}
		\alpha = - \frac{1}{({\cal P}^1)^2 + ({\cal Q}_1)^2}~.
	\end{equation}
We can now diagonalize the mass matrix such that
    \begin{equation} \label{eq:diagonalM}
		\bar{\square} \vec{\mathfrak{Z}} - M^2 \vec{\mathfrak{Z}} = 0~,
    \end{equation}
where $M^2$ is the matrix with the mass eigenvalues of $\mathfrak{M}^2$, and $\vec{\mathfrak{Z}}$ are its eigenvectors. From the definition of the BPS branch, it is clear that all non-diagonal matrix elements \eqref{eq:nondiagonalUngauged} vanish and the mass matrix is automatically diagonal. In the non-BPS branch, $\mathfrak{M}^2$ also simplifies further: depending on the precise distribution of minus signs, $m_{15}$ to $m_{35}$ evaluate to 
    \begin{equation}
        \pm 2\alpha {\cal P}^1{\cal Q}_1 \quad \text{ or } \quad 0~,
    \end{equation}
and $m_{45}, m_{46}$ and $m_{56}$ evaluate to 
    \begin{equation}
        \pm \alpha \left( ({\cal P}^1)^2 - ({\cal Q}_1)^2 \right) \quad \text{ or } \quad 0~.
    \end{equation}
Hence, depending on the assignment of minus signs in \eqref{eq:BPSequalPQs}, the eigenvalues of $\mathfrak{M}^2$ are 
\begin{align} \label{eq:massBPS}
{\rm \bf BPS:}&\qquad  M^2={\rm diag}\le( \frac{2}{\ell_2^2},\cdots,\frac{2}{\ell_2^2}\ri)~, \\ \label{eq:massNonBPS} 
{\rm \bf non-BPS:}& \qquad M^2={\rm diag}\le(0,0,\frac{2}{\ell_2^2},\frac{2}{\ell_2^2},\frac{2}{\ell_2^2},\frac{6}{\ell_2^2} \ri) ~.
\end{align}
Although the mass matrix is more complicated for the non-BPS solutions that do not obey \eqref{eq:BPSequalPQs} but instead have e.g.\ a solution to \eqref{eq:atteqnsg0} with all ${\cal P}^I$ and ${\cal Q}_I$ different, we confirmed numerically that the eigenvalues are still given by \eqref{eq:massNonBPS}, and also for other non-BPS solutions. See App.\,\ref{app:scalings} for details. For this reason, we expect that   \eqref{eq:massNonBPS} captures the spectrum of fluctuations for any non-BPS configuration in STU models. 

In the BPS branch, the eigenvectors are simply the scalar fields themselves. In the non-BPS branch, the eigenvectors are linear combinations of the fields $ (\bph_i, \bch_i)$. 
As an illustrative example, for the case \eqref{eq:nonBPSequalPQs1}, the six eigenstates $\vec{\mathfrak{Z}}=(\mathfrak{Z}_1,\ldots,\mathfrak{Z}_6)$ are
    \begin{equation} \label{eq:nonBPSEV}
    \begin{aligned}
      m^2 \ell_2^2 &= 0~ (\Delta = 1)~: \quad
      \begin{array}{l} 
      {\mathfrak{Z}}_1 = \bch_1 + \bch_3 ~, \\
     \displaystyle {\mathfrak{Z}}_2 = - \frac{2 {\cal P}^1 {\cal Q}_1}{({\cal P}^1)^2 + ({\cal Q}_1)^2} \bph_2 +  \bch_1 + \frac{({\cal P}^1)^2 - ({\cal Q}_1)^2}{({\cal P}^1)^2 + ({\cal Q}_1)^2}\bch_2 ~,
      \end{array} \\[0.5em]
      m^2 \ell_2^2 &= 2~(\Delta = 2)~: \quad 
      \begin{array}{l} 
      {\mathfrak{Z}}_3 = \bph_1~,\\  
      {\mathfrak{Z}}_4 = \bph_3~, \\ 
      \displaystyle {\mathfrak{Z}}_5 =  \le(({\cal P}^1)^2 - ({\cal Q}_1)^2\ri) \bph_2 +  2{\cal P}^1{\cal Q}_1\,\bch_2 ~,
      \end{array}
      \\[0.6em]
      m^2 \ell_2^2 &= 6~(\Delta = 3)~: \quad 
        \begin{array}{l} 
      \displaystyle  {\mathfrak{Z}}_6 = - \frac{2 {\cal P}^1 {\cal Q}_1}{({\cal P}^1)^2 + ({\cal Q}_1)^2} \bph_2 -  \bch_1 + \frac{({\cal P}^1)^2 - ({\cal Q}_1)^2}{({\cal P}^1)^2 + ({\cal Q}_1)^2} \bch_2 +  \bch_3~.
      \end{array}
    \end{aligned}
    \end{equation}
To avoid clutter, here we wrote  eigenstates $\mathfrak{Z}_i$ that are not orthogonal and  we have not included the appropriate normalization to comply with \eqref{eq:kinActiongauged-diag}. When discussing interactions in Sec.\,\ref{sec:ungaugedInt}, we will be referring to a basis that is orthogonal and normalized correctly. We also note that this spectrum is in perfect agreement with the non-BPS branch spectrum analysed in \cite{Larsen:2018cts}.

It is interesting to connect the spectrum of the non-BPS branch to the five-dimensional analysis in \cite{Castro:2018ffi}. When reducing the ungauged five-dimensional theory along a circle, the Myers-Perry black hole can be viewed as an electrically charged solution; its D-brane construction corresponds to the {D0+D6} system, as first noted in \cite{Itzhaki:1998ka}. In the context of the spectrum, we note that the state with $\Delta =3$ here is precisely  the squashing mode $\chi$ in \cite{Castro:2018ffi}.

\subsection{nAttractor revisited}\label{sec:nAtt}

It is interesting to place our linear analysis in the context of  the nAttractor mechanism proposed in \cite{Larsen:2018iou}. The observation  of the nAttractor is that generic scalar moduli, in our case $(\varphi_i,\chi_i)$, should respond in the near-AdS$_2$ region as
\be\label{eq:n-att}
\begin{aligned}
\begin{array}{l}
\displaystyle\varphi_i = \bar \varphi_i + {\varphi_i^{(1)}\over z} +\cdots~, \\[0.6em]  
\displaystyle\chi_i = \bar \chi_i + {\chi_i^{(1)}\over z}+\cdots ~, 
\end{array}
\qquad {\rm as} \quad z\to 0~,
\end{aligned}
\ee
where $ (\bar \varphi_i,  \bar \chi_i)$ are the attractor values of the scalars, and $(\varphi_i^{(1)},\chi_i^{(1)})$ are constants. The power in the AdS$_2$ radial direction $z$ was used as evidence to argue that each of the scalar moduli is dual to an irrelevant operator of conformal dimension $\Delta=2$. 

For the fluctuations around {\bf BPS} backgrounds, this is indeed true: we showed explicitly that the dilaton and axion perturbations, $\bph_i$ and $\bch_i$, are eigenstates  with masses $m^2 = \frac{2}{\ell_2^2}$ and corresponding conformal dimensions $\Delta = 2$.\footnote{This also agrees with the lowest modes in the spectrum of fluctuations of the scalar moduli in \cite{Banerjee:2011jp,Sen:2011ba,Keeler:2014bra}.} Therefore, they have a response as \eqref{eq:n-att} as one moves away from the AdS$_2$ background, in perfect agreement with the nAttractor. Despite sharing the same conformal dimension as $\Y$, it should be noted that  these modes stand on a different footing: only $\Y$ obeys the constrained JT equation \eqref{eq:jt1} which ties its response to thermal AdS$_2$ and the Schwarzian action.  

For the {\bf non-BPS} solutions, the relation to the nAttractor is not as simple. In this case we found that the spectrum of fluctuations contains mass eigenstates with $m^2 = 0$, $m^2 = \frac{2}{\ell_2^2}$, and   $m^2 = \frac{6}{\ell_2^2}$, corresponding to a conformal dimension $\Delta = 1$, $\Delta = 2$ or $\Delta = 3$, respectively. This implies that the eigenstates that are unique to the non-BPS branch behave as
    \begin{equation} \label{eq:fraknonBPS}
    \begin{aligned}
      m^2 \ell_2^2 &= 0~ (\Delta = 1)~: \quad
      {\mathfrak{Z}}_i = {\mathfrak{Z}_i^{(1)}}+O( z^{-2}) 
       \\
      m^2 \ell_2^2 &= 6~(\Delta = 3)~: \quad 
      \displaystyle  {\mathfrak{Z}}_i = {\mathfrak{Z}_i^{(1)}\over z^2}+\cdots~.
    \end{aligned}
    \end{equation}
Hence the marginal operator ($\Delta=0$) could modify the attractor value, and the irrelevant operator with $\Delta=6$ has a different imprint relative to $\Y$.

Two comments are in order regarding the non-BPS branch:
\begin{itemize}
    \item The behaviour \eqref{eq:n-att} persists for non-BPS black holes, as can be seen explicitly from the solutions in \cite{Chow:2014cca}. However, this is not enough to establish that the eigenstates are dual to operators with $\Delta=2$. The reason is that $\varphi_i^{(1)}$ and $\chi_i^{(1)}$ are not independent parameters and cancellations occur. In App.\,\ref{app:scalings}, we set up the extremal and near-horizon limit for the solutions in \cite{Chow:2014cca}: we show how  combinations of $\varphi_i^{(1)}$ and $\chi_i^{(1)}$ cancel for non-BPS black holes and lead to the behaviour in \eqref{eq:fraknonBPS}. 
    \item Having a marginal deformation means that there are flat directions in the attractor mechanism for non-BPS black holes. Concretely, the linear combinations of the attractor values of the scalars corresponding to the eigenstates with $m^2 = 0$ will not be fixed, but can instead take on different (constant) values. The eigenstates with $m^2 \neq 0$ will dictate which linear combinations cannot change value. In the context of non-supersymmetric attractors, the occurrence of flat directions was reported initially in \cite{Tripathy:2005qp}, and it could lead to potential instabilities of the system \cite{Nampuri:2007gv}. 
\end{itemize}

Finally, it is appropriate to add a comment about the gauged solutions and their behaviour in the context of the nAttractor. In that case we can already see in a simple manner  that \eqref{eq:n-att} applies to the moduli, but for slightly different reasons. In the analysis of linear fluctuations in Sec.\,\ref{sec:lineargen}, we have an inhomogeneous solution for the scalar moduli in \eqref{eq:hominh}, which implies that the moduli behave in the near-AdS$_2$ region as
 \be
 (\varphi_i, e^{\bar\varphi_i} \chi_i)=(\bar\varphi_i, e^{\bar\varphi_i} \bar\chi_i)+  \vec{a}  {\Y\over \Phi_0} + (\bph_i, e^{\bar\varphi_i} \bch_i)~,
 \ee
with the second term complying with \eqref{eq:n-att}. This is also evident if one inspects non-extremal black holes in  gauged supergravity.\footnote{A straightforward check can be done with, e.g., the solutions in \cite{Cacciatori:2009iz,Chow:2013gba,Benini:2015eyy}.}
But in this case it is also clear that the moduli do not have $\Delta=2$: there is just a mixing of the JT field with the matter sector that needs to be diagonalized. The operator interpretation of the moduli comes from $(\bph_i, \bch_i)$ which contain the independent degrees of freedom. 
 
\subsection{Interactions} \label{sec:ungaugedInt}
 
The classification of the interactions is relatively simple in the ungauged case, since the JT sector does not mix with the matter fields at the linear level. The effective Euclidean action for the matter fields around the near-AdS$_2$ background is of the form
  \be\label{eq:effUG}
 \begin{aligned}
 I_{\rm eff} = \int d^2 x \sqrt{-\bar g^{(2)}} \,\le({\cal L}_{\rm kin}+ {\cal L}_{\rm int-\Y}\ri)~,
  \end{aligned}
 \ee
The quadratic terms for the scalar fields, which contain the kinetic and mass terms, are
 \be \label{eq:kinActionUngauged}
  \begin{aligned}
 {\cal L}_{\rm kin}=\frac{1}{2} \partial_a \vec{\mathfrak{Z}} \cdot \partial^a \vec{\mathfrak{Z}} +\frac{1}{2} \vec{\mathfrak{Z}}^T {M^2} \vec{\mathfrak{Z}} ~.
\end{aligned}
 \ee
Here  $\vec{\mathfrak{Z}} $ contains the degrees of freedom for the supergravity fields $(\varphi_i,\chi_i)$ and are orthogonal at leading order in the near-AdS$_2$ region; $M^2$ is the matrix with the mass eigenvalues as defined in \eqref{eq:diagonalM}. We are interested in finding the corrections to the two-point functions of the fields $\vec{\mathfrak{Z}}$ due to the interactions with $\Y$ discussed in \eqref{eq:kinActiongauged-diag}. The terms in the effective action that involve cubic interactions with one power of $\Y$ are very simple for the ungauged theory, and read
 \be \label{eq:intActionUngauged}
 \begin{aligned}
 {\cal L}_{\rm int}= \frac{1}{\Phi_0}{ \Y } \partial_a \vec{\mathfrak{Z}} \cdot \partial^a \vec{\mathfrak{Z}} - \frac{3}{2\Phi_0}\Y {\vec{\mathfrak{Z}}}^T {M^2} \vec{\mathfrak{Z}} ~.
\end{aligned}
 \ee
As we discussed in Sec.\,\ref{sec:interact}, we want to report on how these interactions affect the two-point functions of $\vec{\mathfrak{Z}}$. The final answer is of the form
    \begin{equation}
        \langle\mathcal{O}_{\mathfrak{Z}_i}(u_1)\mathcal{O}_{\mathfrak{Z}_i}(u_2)\rangle_{\beta}=\left[\frac{\pi}{\beta\sin(\frac{\pi u_{12}}{\beta})}\right]^{2\Delta}\left[D+\frac{\tilde{D} a\beta^2}{2\pi^2}\left(2+\pi\frac{1-2u_{12}/\beta}{\tan(\frac{\pi u_{12}}{\beta})}\right)\right]~.
    \end{equation}
Following from \eqref{defTilD}, the parameter $\tilde{D}$  for this case is 
    \begin{equation} \label{eq:defDtilde}
        \tilde{D} = \lambda_{\Y \mathfrak{Z}_i \mathfrak{Z}_i} K_{\Y \mathfrak{Z}_i \mathfrak{Z}_i} + {\lambda}_{\Y (\partial\mathfrak{Z}_i) (\partial\mathfrak{Z}_i)} {K}_{\Y (\partial\mathfrak{Z}_i) (\partial\mathfrak{Z}_i)}~, 
    \end{equation}
where the couplings $\lambda_{\Y \mathfrak{Z}_i \mathfrak{Z}_i}$ and ${\lambda}_{\Y (\partial\mathfrak{Z}_i) (\partial\mathfrak{Z}_i)}$ can be read off from the interaction term in the cubic action, and $K_{\Y \mathfrak{Z}_i \mathfrak{Z}_i}$ and ${K}_{\Y (\partial\mathfrak{Z}_i) (\partial\mathfrak{Z}_i)}$ appear in the three-point functions of the operators dual to the fields $\vec{\mathfrak{Z}}$ and $\Y$; they are functions of the conformal dimensions of these operators. 
From \eqref{eq:intActionUngauged}, we have
    \begin{equation}
        \lambda_{\Y \mathfrak{Z}_i \mathfrak{Z}_i} = -\frac{3}{2} m_i^2~, \qquad {\lambda}_{\Y (\partial\mathfrak{Z}_i) (\partial\mathfrak{Z}_i)} = 1~, 
    \end{equation}
$K_{\Y \mathfrak{Z}_i \mathfrak{Z}_i}$ is given by \eqref{Kcoeff} and
    \begin{equation}
    {K}_{\Y (\partial\mathfrak{Z}_i) (\partial\mathfrak{Z}_i)} = - \frac{\Delta^2 - \Delta - 1}{\ell_2^2} K_{\Y \mathfrak{Z}_i \mathfrak{Z}_i}~,
    \end{equation}
such that
    \begin{equation} \label{eq:Dtilde}
        \tilde{D}_i =- \frac{1}{\ell_2^2} \left( \frac{5}{2} \Delta(\Delta - 1) -1 \right)K_{\Y \mathfrak{Z}_i \mathfrak{Z}_i}~.
    \end{equation}
We will now report on the correction to the two-point functions due to the interaction with $\Y$ for both the BPS and the non-BPS branch. 

\paragraph{BPS branch.}   Here, $\vec{\mathfrak{Z}} = (\bph_i, \bch_i)$ are eigenstates of the mass matrix, and all eigenvalues of $M^2$ are $m^2 = \frac{2}{\ell_2^2}$. 
For these fields, $\Delta_{\mathfrak{Z}} = 2 $, such that \eqref{eq:defD} and \eqref{eq:Dtilde} give, respectively,
    \begin{equation} \label{eq:Dbps}
        D = \frac{6}{\pi}~, \qquad       \tilde{D}  = \frac{18}{\pi\ell_2^2}~,
    \end{equation}
where we used $K_{\Y \mathfrak{Z}_i \mathfrak{Z}_i}=-{9/2\pi}$.

\paragraph{Non-BPS branch.} The structure of the corrected two-point function is similar to the one in the BPS branch, but there are important differences. The main differences come from the spectrum of conformal dimensions we have in this branch, listed in \eqref{eq:massBPS}. We have three eigenstates with mass $m^2 = \frac{2}{\ell_2^2}$ and $\Delta = 2$; for these fields, the values of $D$ and $\tilde{D}$ are equivalent to those in the BPS branch in \eqref{eq:Dbps}. For the eigenstate with mass $m^2= \frac{6}{\ell_2^2}$ and $\Delta = 3$, one obtains
    \begin{equation}\label{eq:tD6mass}
        D = \frac{40}{3\pi}~, \qquad \tilde{D} = \frac{70}{\pi\ell_2^2} ~,
    \end{equation}
with $K_{\Y \mathfrak{Z}_i \mathfrak{Z}_i}= -{5/\pi}$.

However, the  two eigenstates  with mass $m^2 = 0$, and corresponding conformal dimension $\Delta = 1 $, are more delicate. 
As is clear from \eqref{Kcoeff}, at $\Delta_{\mathfrak{Z}} = 1$ the coefficient $K_{\Y \mathfrak{Z} \mathfrak{Z}}$ diverges. This divergence originates from the fact that the conformal dimensions of the operators considered sum to $\Delta_\Y + 2 \Delta_{\mathfrak{Z}}=d$, with $\Delta_{\Y} = -1$, and $d=1$. This is a known divergence that is related to extremal correlators \cite{DHoker:1999jke}.\footnote{In an extremal three-point function, the conformal dimension of one of the operators is equal to the sum of the remaining ones, e.g.\  $\Delta_1 = \Delta_2 + \Delta_3$. For the evaluation of \eqref{eq:threep}, the irrelevant deformation $\Y$ is in the $\Delta_{-}$ branch, such that we have $d = \Delta_-+ \Delta_2 + \Delta_3$ which is the same statement.} Since the cubic coupling in \eqref{eq:intActionUngauged} is clearly not zero, we find that the corresponding $\tilde{D}$ is divergent.

The appearance of an extremal correlator with a non-zero cubic coupling is unusual and problematic.   Unusual, because in higher-dimensional theories of AdS$_{d+1}$ --- arising from a consistent truncation of a ten-dimensional compactification --- the extremal cubic couplings are zero; for theories with an AdS$_5$ factor see \cite{DHoker:1999jke,Lee:1998bxa,Liu:1999kg}, and \cite{Mihailescu:1999cj,Arutyunov:2000by,Taylor:2007hs} for cases in AdS$_3$. Problematic, because the divergence forces the  conclusion that the non-BPS branch does not lead to a near-AdS$_2$ background with well-defined correlators.

There are two additional comments in this regard:
\begin{itemize}
    \item Let us first consider the possibility that \eqref{eq:intActionUngauged} is missing terms, and if included appropriately they should lead to a vanishing cubic coupling of the marginal operator and $\Y$. In particular we will include $h_{ab}$ (assuming it is has a background value controlled by $\Y$). The presence of $h_{ab}$ adds the following modification to \eqref{eq:intActionUngauged} (up to overall normalizations)
    \be
 h^{\textrm{ST}\,ab}\partial_a \vec{\mathfrak{Z}} \cdot \partial_b \vec{\mathfrak{Z}}  -   \frac{1}{2} \hat{h}  \vec{\mathfrak{Z}}^T {M^2} \vec{\mathfrak{Z}}   ~.
    \ee
    The trace mode, $\hat h$, does not contribute for a marginal (massless) mode, and moreover it is also decoupled from $\Y$ as reflected in \eqref{eq:traceeomh}. For the symmetric traceless piece, we note that
\begin{equation}
    \begin{aligned}
         h^{\textrm{ST}\,ab}\partial_a \vec{\mathfrak{Z}} \cdot \partial_b \vec{\mathfrak{Z}} &= \bar\nabla^a \bar\nabla^b U(x) \left(\partial_a \vec{\mathfrak{Z}} \cdot \partial_b \vec{\mathfrak{Z}}  -\frac{1}{2} g_{ab}\partial_c \vec{\mathfrak{Z}} \cdot \partial^c \vec{\mathfrak{Z}}\right)\\
         &=  - \bar\nabla^b U(x) (\bar \square \vec{\mathfrak{Z}}) \cdot \partial_b \vec{\mathfrak{Z}} + ({\rm total ~derivative})
    \end{aligned}
\end{equation}    
which is again zero for the massless mode. This shows that the states with $\Delta=1$ do not couple to $h_{ab}$, and hence would not affect the cubic coupling with $\Y$ even if we make $h_{ab}$ depend on the JT field.
    \item In the non-BPS sector there are other extremal correlators: cubic couplings between the moduli such that  $\Delta_i = \Delta_j + \Delta_k$, which is a common occurrence since we have operators with $\Delta=1,2,3$. A simple computation for our theory shows that all of these couplings $\lambda_{ijk}$ are zero. This confirms that within a consistent supergravity truncation the couplings vanish as in the higher-dimensional AdS cases.  
\end{itemize}
Based on this, the conclusion we reach is that for the non-BPS branch we have not identified correctly the near-AdS$_2$ background that describes the backreaction as we turn on the irrelevant deformation $\Y$. Unfortunately it is not clear to us how to modify our definitions and setup to fix this problem, while keeping the marginal operator as part of the spectrum.  

\section{Examples in gauged supergravity}\label{sec:gauged-cases} 

In this section we will consider some special cases in the gauged theory. In Sec.\,\ref{sec:magneticnonBPS} and \ref{sec:magneticBPS} we restrict to purely magnetic solutions, i.e.\ $Q_I = 0$ and $P^I \neq 0$, and consider two subsets of solutions to the attractor equations: non-BPS solutions following the conventions of \cite{Chow:2013gba}, and BPS solutions described in \cite{Cacciatori:2009iz,Benini:2015eyy}. We collect the attractor equations for both cases in App.\,\ref{app:ads2}, and below discuss the non-universal properties that these backgrounds dictate at the level of the spectrum of operators and the interactions. Finally, in Sec.\,\ref{sec:4DBH} we briefly consider a simple dyonic non-BPS example.

\subsection{Magnetic non-BPS background} \label{sec:magneticnonBPS}
Here we will analyse a non-BPS background in the gauged theory that is supported by magnetic charges. It is a specific background that has a well-defined limit as we take $g\to0$, and hence connects with the ungauged backgrounds. Since it is cumbersome to keep four independent charges $P^I$, we will further specialize to the case where $P^1 = P^2 = P^3$, and $P^4$ is independent; appendices \ref{app:ads2} and \ref{app:magnetic} contain expressions when all four charges are independent. 

Although we will have only two independent charges, we will see below that this example already captures interesting non-universal features. Some of these features depend on the relative sign of $P^1$ and $P^4$; to simplify our discussion and make the analysis more transparent, we will sometimes set $P^4 = \pm P^1$, which corresponds to the near-AdS$_2$ region of the magnetic RN black hole in AdS$_4$. 

\paragraph{\texorpdfstring{AdS$_2$}{AdS2} background.} For the attractor values of the scalars, the above simplification means
    \begin{equation}
        \barp_1 = \barp_2 = \barp_3~, 
    \end{equation}
and the remaining background equations fully determine the remaining fields $\barp_1, \Phi_0$ and the AdS$_2$ radius $\ell_2$ via
    \begin{equation} \label{eq:3equalPsatt}
    \begin{aligned}
        4g^2\Phi_0^4 \sinh\barp_1 &= e^{\barp_1}(P^1)^2 - e^{-3\barp_1}(P^4)^2~, \\
        4g^2\Phi_0^4 \cosh\barp_1 &= e^{\barp_1}(P^1)^2 + \frac{1}{3} e^{-3\barp_1}(P^4)^2 - \frac{4\Phi_0^2}{3}~, \\
        \frac{1}{\ell_2^2} &= \frac{1}{\Phi_0^2} + 6g^2 \cosh\barp_1 ~.
    \end{aligned}
    \end{equation}
If $P^4 = \pm P^1$, the attractor equations simplify: this sets $\barp_1 = 0$, and \eqref{eq:3equalPsatt} reduces to
    \begin{equation} \label{eq:simpleatteqs}
    \begin{aligned}
        \frac{1}{\ell_2^2} - \frac{1}{\Phi_0^2} &= 6g^2~, \\
        \frac{1}{\ell_2^2} + \frac{1}{\Phi_0^2} &= \frac{2(P^1)^2}{\Phi_0^4}~.
    \end{aligned}
    \end{equation}

\paragraph{Spectrum of operators.} Diagonalizing the mass matrix for both sectors as 
    \begin{equation}
        \bar{\square} \vec{\mathfrak{Z}} - M^2 \vec{\mathfrak{Z}} = 0~,
    \end{equation}
we find its eigenvalues $m^2$ and corresponding orthogonal eigenstates $\vec{\mathfrak{Z}} = (\mathfrak{Z}_1, \ldots, \mathfrak{Z}_6)$ to be 
    \begin{equation} \label{eq:mass3equalnonBPS}
    \begin{aligned}
      m_1^2  &= \frac{1}{\Phi_0^2} + e^{-\barp_1} \left(g^2 + \frac{P^1 P^4 }{\Phi_0^4} \right)~: 
      &&\qquad \begin{array}{l} 
      {\mathfrak{Z}}_1 = -\bch_1^{\rm hom} + \bch_3^{\rm hom} ~, \\
        {\mathfrak{Z}}_2 = -\bch_1^{\rm hom} + 2\bch_2^{\rm hom} -\bch_3^{\rm hom}~,
      \end{array} \\[0.5em]
      m_2^2  &= \frac{2}{\Phi_0^2} + 4g^2 e^{-\barp_1}~: 
      && \qquad \begin{array}{l} 
      \displaystyle {\mathfrak{Z}}_3 = \bph_1^{\rm hom} + \bph_2^{\rm hom} + \bph_3^{\rm hom} ~, 
      \end{array} \\[0.5em]
      m_3^2  &= \frac{2}{\Phi_0^2} + g^2(3e^{\barp_1} + e^{-\barp_1})~: 
      &&\qquad \begin{array}{l} 
      {\mathfrak{Z}}_4 = -\bph_1^{\rm hom} + \bph_3^{\rm hom} ~, \\
        {\mathfrak{Z}}_5 = -\bph_1^{\rm hom} + 2\bph_2^{\rm hom} - \bph_3^{\rm hom} ~,
      \end{array}
      \\[0.6em]
      m_4^2  &= \frac{2}{3\Phi_0^2} + \frac{e^{-3\barp_1}(P^4 - 3e^{2\barp_1}P^1)^2}{3\Phi_0^4}~: 
      &&\qquad \begin{array}{l} 
      {\mathfrak{Z}}_6 = \bch_1^{\rm hom} + \bch_2^{\rm hom} + \bch_3^{\rm hom} ~. 
      \end{array}
    \end{aligned}
    \end{equation}
If $P^4 = P^1$, there is only one degenerate mass: 
    \begin{equation} \label{eq:P14samesign}
        m^2 = 4g^2 + \frac{2}{\Phi_0^2}~: \qquad \vec{\mathfrak{Z}} = (\bph_i, \bch_i)~.
    \end{equation}
If $P^4 = - P^1$, we have 
    \begin{equation} \label{eq:P14oppossign}
    \begin{aligned}
      &m_1^2  = -2g^2~: 
      &&\qquad \begin{array}{l} 
      {\mathfrak{Z}}_1 = -\bch_1^{\rm hom} + \bch_3^{\rm hom} ~, \\
        {\mathfrak{Z}}_2 = -\bch_1^{\rm hom} + 2\bch_2^{\rm hom} -\bch_3^{\rm hom}~,
      \end{array} \\[0.5em]
      &m_{2,3}^2 = m_\varphi^2 = 4g^2 + \frac{2}{\Phi_0^2}~: 
      && \qquad \begin{array}{l} 
      \displaystyle {\mathfrak{Z}}_{i + 2} = \bph_i ~, 
      \end{array} \\[0.5em]
      &m_4^2  = \frac{2}{3}\left( \frac{4}{\ell_2^2} + \frac{5}{\Phi_0^2}\right)~: 
      &&\qquad \begin{array}{l} 
      {\mathfrak{Z}}_6 = \bch_1^{\rm hom} + \bch_2^{\rm hom} + \bch_3^{\rm hom} ~. 
      \end{array}
    \end{aligned}
    \end{equation}
    
It is instructive to inspect the conformal dimensions associated to the states \eqref{eq:mass3equalnonBPS} in more detail. Starting with the second one listed in \eqref{eq:mass3equalnonBPS} we have
    \be
    \begin{aligned}\label{eq:b1}
\Delta_2&= \frac{1}{2}+\sqrt{\frac{1}{4}+m_1^2\ell_2^2}\\
&= \frac{1}{2}+\sqrt{\frac{9}{4}-2g^2\ell_2^2 e^{-\barp_1}-6g^2\ell_2^2 e^{\barp_1} }~.
    \end{aligned}
    \ee
It is interesting to note that the effect of the AdS$_4$ surrounding ($g\neq0$) is that it lowers the conformal dimension relative to the ungauged case ($g=0$) where the corresponding state has $\Delta_2=2$.
From \eqref{eq:3equalPsatt}, the lower bound of $\Delta_2$ is attained when $\bar\varphi_1=0$ (i.e.\ when $P^1 = \pm P^4$) and  $6g^2\ell_2^2<1$. This implies that the range of $\Delta_2$ is
    \be \label{eq:rangeDelta1}
1.46 < \Delta_2 \leq 2~,
    \ee
hence making it an irrelevant operator below the JT field $\Y$. A similar analysis will show the same range of values applies for $\Delta_3$. For the case $P^4 = P^1$, all conformal dimensions have the range \eqref{eq:rangeDelta1}, and for $P^4 = - P^1$, it applies to the three eigenstates with $\Delta_\varphi$. 
    
For $m_1^2$ and $m_4^2$ the analysis is more delicate: their final values depend on the choice of signs of $P^1$ and $P^4$ and cannot solely be determined from \eqref{eq:3equalPsatt}. If $P^1$ and $P^4$ have the same sign, a similar analysis will show that the ranges of $\Delta_1$ and $\Delta_4$ are the same as \eqref{eq:rangeDelta1}. In particular, this clearly holds for the case $P^4 = P^1$ given in \eqref{eq:P14samesign}. However, if $P^1$ and $P^4$ have opposite sign, we find that both bounds on $\Delta_4$ are increased: 
    \begin{equation}
        2.21 < \Delta_4 \leq 3~,
    \end{equation}
such that now $\Delta_4 > \Delta_\Y$. For this choice of sign, the mass $m_1^2$ in \eqref{eq:mass3equalnonBPS} can become negative; hence, it can violate the Breitenlohner-Freedman bound. In particular, for $P^4 = - P^1$ it is clear from \eqref{eq:P14oppossign} that the mass $m_1^2 <0$, 
and $\Delta_1$ reduces to
    \begin{equation}
        \Delta_1 = \frac{1}{2} + \frac{1}{2}\sqrt{1-8g^2\ell_2^2}~.
    \end{equation}
Thus, when $P^1 = - P^4$, the eigenstate with mass-squared $m_1^2$ violates the BF bound if 
    \begin{equation} \label{eq:BFviolation}
        8g^2\ell_2^2 < 1~.
    \end{equation}
This is a stricter requirement than the one implied by \eqref{eq:simpleatteqs}, i.e.\ $6g^2\ell_2^2 < 1$. In Sec.\,\ref{sec:4DBH} we discuss a dyonic example, $P^1 = \pm Q_1$ and the remaining charges equal to zero, for which we find the same bound \eqref{eq:BFviolation}. 

Finally, notice that we can smoothly take the $g \to 0$ limit in \eqref{eq:3equalPsatt} and \eqref{eq:mass3equalnonBPS}. 
The remaining attractor equations \eqref{eq:3equalPsatt} then determine the charges up to a relative sign between $P^4$ and $P^1$. The three eigenvalues in \eqref{eq:mass3equalnonBPS} corresponding to $\mathfrak{Z}_{3,4,5}$ collapse in either case to $m^2 \ell_2^2 = 2$; for the remaining three this depends on this relative sign. If $P^1$ and $P^4$ have the same sign, the six eigenvalues are as in the ungauged BPS case \eqref{eq:massBPS}; if there is a relative sign, we instead land in the non-BPS case \eqref{eq:massNonBPS}.

\paragraph{Interactions.} We are now ready to discuss the interactions in the magnetic non-BPS case introduced above. The general expressions are presented in App.\,\ref{app:magnetic}. For clarity here we present the values for the limiting case where all magnetic charges are equal in size (i.e.\ we further set $P^1 = \pm P^4$). 

As in Sec.\ \ref{sec:ungaugedInt}, we will classify the corrections to the two-point functions in terms of the parameter $\hat{D}$ defined in \eqref{eq:Dhat}. If $P^1= P^4$, all $\hat D_i$ are equal, and their value is
\begin{equation}
    \begin{aligned}
        \hat D_i&= - \frac{3}{4(\Delta_i - 1)}\le(-\frac{4}{\ell_2^2}+ 16 g^2\ri)~.
    \end{aligned}
\end{equation}
Also, all conformal dimensions are equal to \eqref{eq:b1} with $\bar\varphi_1=0$. Using this in \eqref{Kcoeff} to compute $K_{\Y \mathfrak{Z}_i \mathfrak{Z}_i}$ then determines the range of $\hat D_i$: as $0 \leq g^2 < \frac{1}{6\ell_2^2}$,
    \begin{equation} \label{eq:equaltD1}
        2.19 < \ell_2^2 \hat D_i \leq  3~.
    \end{equation}
Thus, $\hat{D}_i$ is always positive and setting $g=0$ here agrees precisely with the BPS case \eqref{eq:Dbps}, as expected. 

If instead $P^1= -P^4$, there are different behaviours since the eigenvalues do not all become degenerate. We obtain 
\begin{equation}
    \begin{aligned} \label{eq:oppositeDtildes}
        \hat D_1&= - \frac{3}{4(\Delta_1 - 1)} \le(\frac{1}{\ell_2^2}+ g^2\ri)~, \\
        \hat D_\varphi &= - \frac{3}{4(\Delta_\varphi - 1)}\le(-\frac{4}{\ell_2^2}+ 16 g^2\ri)~,\\
        \hat D_4&= - \frac{3}{4(\Delta_4 - 1)}\le(-\frac{14}{\ell_2^2}+ 46 g^2\ri)~.
    \end{aligned}
\end{equation}
The range of $\hat D_\varphi$ is as in \eqref{eq:equaltD1}, and for $\hat D_4$ we find 
    \begin{equation}
        3.93  < \ell_2^2 \hat D_4 \leq \frac{21}{4}~,
    \end{equation}
i.e.\ $\hat D_4$ is always positive as well. Setting $g=0$ matches perfectly with the non-BPS case: $\hat D_\varphi$ matches with \eqref{eq:Dbps} and $\hat D_4$ with \eqref{eq:tD6mass}. The case of $\hat D_1$ is more subtle: for $g = 0$ we get $\Delta_1 = 1$ and the correlator is extremal; and it also violates the BF bound for $g^2 > \frac{1}{8\ell_2^2}$. Within the range $g^2 \in (0, \frac{1}{8\ell_2^2})$, however, $\hat{D}_1$ is positive. 

\subsection{Magnetic BPS background} \label{sec:magneticBPS}
The solution in Sec.\,\ref{sec:magneticnonBPS} is related to the BPS solution of \cite{Benini:2015eyy}, see also \cite{Cacciatori:2009iz,Hristov:2010ri}. In App.\,\ref{app:ads2} we relate the attractor solution of \cite{Benini:2015eyy} to the notation used in the previous section. An important point is that the BPS solution \eqref{eq:cbps2} is at a very different footing as compared to the non-BPS solution \eqref{eq:magneticAttractor}: all charges are proportional to $g$, so they are not smoothly connected to solutions in the ungauged theory. 

For ease of comparison with the literature, we will adopt the conventions of \cite{Hristov:2010ri,Benini:2015eyy} to describe the magnetic charges. Hence we will use
\be \label{eq:ntoP1}
\mathfrak{n}_1= g P^4~, \quad \mathfrak{n}_2 = g P^2~, \quad \mathfrak{n}_3 = g P^3~, \quad \mathfrak{n}_4 = g P^1~,
\ee
where $\mathfrak{n}_I$ are integral charges.  

\paragraph{AdS$_2$ background.} We will again consider a subset of solutions for which three charges are equal and negative, and the remaining charge is positive. To comply with the conditions on the charges $\mathfrak{n}_i$ given in \eqref{eq:condsum}, we choose  $\mathfrak{n}_2 = \mathfrak{n}_3 = \mathfrak{n}_4 < 0$ and $\mathfrak{n}_1 > 0$ such that
    \begin{equation}
        \mathfrak{n}_1 + 3 \mathfrak{n}_4 = 2~.
    \end{equation}
Thus, contrary to the non-BPS case, there is only one independent charge. This sets
    \begin{equation}
        \barp_1 = \barp_2 = \barp_3~.
    \end{equation}
From \eqref{eq:cbps2}, the charges can be solved for as 
    \begin{equation}
       \mathfrak{n}_1 = \frac{g^2\Phi_0^2}{2} \left( e^{3\barp_1} - 3 e^{\barp_1} \right)~,  \quad   \mathfrak{n}_{2,3,4} = - g^2\Phi_0^2 \cosh{\barp_1}~,
    \end{equation}
and \eqref{eq:L2Benini} and \eqref{eq:Phi0Benini} give
    \begin{equation} \label{eq:L2Phi0equalcharges}
        \frac{1}{\ell_2^2} = \frac{g^2}{4} e^{-\barp_1}(3+e^{2\barp_1})^2~, \qquad \frac{1}{\Phi_0^2} = - \frac{2g^2e^{-\barp_1}}{\mathfrak{n}_4 + e^{-2\barp_1} \mathfrak{n}_1}~.
    \end{equation}
In terms of $\barp_1$ and $\Phi_0$, the constraint on the charges gives 
    \begin{equation} \label{eq:condchargesinL}
        \frac{1}{\Phi_0^2} = \frac{g^2}{4}e^{-\barp_1} \left(e^{4\barp_1}-3(2e^{\barp_1}+1)\right) ~.
    \end{equation}
    
\paragraph{Spectrum of operators.} At the linearized level, the inhomogeneous solution for $\bph_i$ is the same as in \eqref{eq:inhomMagnetic}; explicitly, for the special case we consider, we have
    \begin{equation}
        \bph_i = \frac{2}{\Phi_0} \frac{1- e^{2\barp_1}}{1+e^{2\barp_1}} \Y + \bph_i^{\rm hom}~,
    \end{equation}
and the inhomogeneous parts of $\bch_i$ can consistently be set to vanish. Diagonalizing the mass matrix for both sectors $(\bph_i^{\rm hom}, \bch_i^{\rm hom })$ as
    \begin{equation}
        \bar{\square} \vec{\mathfrak{Z}} - M^2 \vec{\mathfrak{Z}} = 0~,
    \end{equation}
we find the mass eigenvalues and corresponding (orthogonal) eigenstates to be:
    \begin{equation}\label{eq:Beninimass}
    \begin{aligned} 
      m_1^2  &= -2g^2\sinh\barp_1~: 
      &&\qquad \begin{array}{l} 
      {\mathfrak{Z}}_1 = -\bch_1^{\rm hom} + \bch_3^{\rm hom} ~, \\
        {\mathfrak{Z}}_2 = -\bch_1^{\rm hom} + 2\bch_2^{\rm hom} - \bch_3^{\rm hom} ~,
      \end{array} \\[0.5em]
      m_2^2  &= \frac{2}{\Phi_0^2} + 4g^2 e^{-\barp_1}~: 
       &&\qquad \begin{array}{l} 
      \displaystyle {\mathfrak{Z}}_3 = \bph_1^{\rm hom} + \bph_2^{\rm hom} + \bph_3^{\rm hom} ~, 
      \end{array} \\[0.5em]
      m_3^2 &= \frac{2}{\Phi_0^2} + g^2 e^{\barp_1}(3 + e^{-2\barp_1}): 
     &&\qquad \begin{array}{l} 
      {\mathfrak{Z}}_4 = -\bph_1^{\rm hom} + \bph_3^{\rm hom} ~, \\
        {\mathfrak{Z}}_5 = -\bph_1^{\rm hom} + 2\bph_2^{\rm hom} - \bph_3^{\rm hom}~,
      \end{array}
      \\[0.6em]
      m_4^2 &= \frac{2}{\Phi_0^2} + 4g^2 e^{\barp_1}\cosh^2\barp_1~: 
      &&\qquad \begin{array}{l} 
      {\mathfrak{Z}}_6 = \bch_1^{\rm hom} + \bch_2^{\rm hom} + \bch_3^{\rm hom} ~, 
      \end{array}
    \end{aligned}
    \end{equation}
We can write the eigenvalues fully in terms of the charge $\mathfrak{n}_4$ and the gauge coupling $g$ using \eqref{eq:condchargesinL} and 
    \begin{equation} \label{eq:L3Benini}
       e^{-2\barp_1} = \frac{\mathfrak{n}_4}{3\mathfrak{n}_4 - 1 - \sqrt{(2\mathfrak{n}_4-1)(6\mathfrak{n}_4-1)}}~.
    \end{equation}
Then it is clear that the eigenvalues are all proportional to $g^2$, as expected. 

Again, it is instructive to inspect the conformal dimensions associated to these eigenstates. Starting with the second listed eigenvalue, in terms of $\mathfrak{n}_4$ only we have
    \begin{equation} \label{eq:delta1BPSmagn}
    \begin{aligned}
        \Delta_2 = 2\sqrt{\frac{1-2\mathfrak{n}_4}{1-6\mathfrak{n}_4
        }} ~.
    \end{aligned}
    \end{equation}
From the conditions on the charges, it is clear that $\mathfrak{n}_4 \in (-\infty, -1)$. This determines the range of $\Delta_2$ as 
    \begin{equation}
        \frac{2}{\sqrt{3}} < \Delta_2 < 2\sqrt{\frac{3}{7}}~.
    \end{equation}
Similarly, we find for $\Delta_3$,
    \begin{equation}
        \Delta_3 = 1 + \sqrt{\frac{1-2\mathfrak{n}_4}{1-6\mathfrak{n}_4
        }}~, \qquad 1+\frac{1}{\sqrt{3}}< \Delta_3 < 1+\sqrt{\frac{3}{7}}~, 
    \end{equation}
and for $\Delta_4$
    \begin{equation}
        \Delta_4 
        =1+2\sqrt{\frac{1-2\mathfrak{n}_4}{1-6\mathfrak{n}_4}}~,
        \qquad 1+\frac{2}{\sqrt{3}} < \Delta_4 < 1+2\sqrt{\frac{3}{7}}~.
    \end{equation}
Thus, comparing to the irrelevant deformation $\Y$ we have $1<\Delta_2 < \Delta_3 < \Delta_{\Y} < \Delta_4<2.5$, so $\mathfrak{Z}_{3,4,5}$ are less and $\mathfrak{Z}_{6}$ is more irrelevant than the JT field $\Y$. Finally, $\Delta_1$ is given by 
    \begin{equation} \label{eq:delta4BPS}
        \Delta_1^{\pm} 
        = \frac{1}{2} \pm \left( -\frac{1}{2} + \sqrt{\frac{1-2\mathfrak{n}_4}{1-6\mathfrak{n}_4}}\right)~, 
    \end{equation}
where   
    \begin{equation} \label{eq:delta4BPS1}
        \frac{1}{\sqrt{3}} < \Delta_1^+ < \sqrt{\frac{3}{7}}~, \qquad 1-\sqrt{\frac{3}{7}} < \Delta^-_1 < 1-\frac{1}{\sqrt{3}}~. 
    \end{equation}
From \eqref{eq:delta4BPS} it is clear that the fourth mass $m_4^2$ takes on negative values. However, the BF bound is never violated, as both $\Delta_1^\pm$ are real for any value of the charge $\mathfrak{n}_4$. The difference between $\Delta_{2,3,4}$ and $\Delta_1$ can also be seen directly from \eqref{eq:Beninimass}. The last three eigenvalues are manifestly positive, but for the first eigenvalue we can consider the BF bound, $m^2 + \frac{1}{4\ell_2^2} \geq 0$. With the AdS$_2$ radius as in \eqref{eq:L2Phi0equalcharges} we find that it is never violated: 
    \begin{equation}
        m_1^2 + \frac{1}{4\ell_2^2} = \frac{g^2}{16}e^{-\barp_1}  (e^{2\barp_1} -5)^2 \geq 0~. 
    \end{equation}
    
\paragraph{Interactions.} Finally, we discuss the corrections to the two-point functions of the eigenstates $\mathfrak{Z}_i$ due to cubic interactions with $\Y$. The corrections are of the form \eqref{two:total}; the general expressions are not very insightful, so we give one explicit example: 
\begin{equation}
    \hat D_2 = - \frac{3}{4(\Delta_2-1)} \frac{g^2 e^{-\barp_1}}{2(1+3e^{2\barp_1})^2} \le( -9 e^{8\barp_1} + 86 e^{6\barp_1} + 44 e^{4\barp_1} - 118 e^{2\barp_1} - 3\ri)~,
\end{equation}
where $\Delta_2$ is given in \eqref{eq:delta1BPSmagn}, and we can use \eqref{eq:L3Benini} to write this in terms of the charge $\mathfrak{n}_4$ only. Then $\mathfrak{n}_4 \in (-\infty, -1)$ determines 
    \begin{equation}
        -19.7\,g^2 < \hat D_2 < -4.49\,g^2~.
    \end{equation}
A similar calculation gives
    \begin{equation}
    \begin{aligned}
        25.3\,g^2 < & \hat D_3 < 34.1\,g^2~, \\
        35.4\,g^2 < & \hat D_4 < 48.7\,g^2~,
    \end{aligned}
    \end{equation}
and 
    \begin{equation}
        10.3\,g^2 < \hat D_1^- < 12.4\,g^2~, \qquad  14.1\,g^2 < \hat D_1^+ < 23.5\,g^2~.
    \end{equation}
Thus only $\hat D_2$ is negative, and all other $\hat D_i$ are positive. The sign of $\hat{D}$ is an interplay of the sign of the effective cubic coupling constant $\lambda_{\rm eff}$ of $\Y$ and the field $\mathfrak{Z}_i$, and of the conformal dimension $\Delta_i$: the prefactor in \eqref{eq:Dhat} is positive for $\Delta <1$, and negative otherwise. This is the only background for which we find a different sign for the corrections; it would be interesting to reproduce this from a near-CFT$_1$ description, or account for it from the dual CFT$_3$ in the UV.

\subsection{Dyonic, non-BPS}\label{sec:4DBH}
 
In this section, we consider an additional example to complement the discussion in Sec.\,\ref{sec:magneticnonBPS} and \ref{sec:magneticBPS}. It is a very simple configuration: all but one field strength vanishes, and the magnetic and electric charge are matched. More concretely, here we will take $P^1=\pm Q_1$ and all remaining charges equal to zero. This example is non-BPS, and hence we will find similar results as those described in Sec.\,\ref{sec:magneticnonBPS}. The corresponding black hole solution can be found in, e.g., \cite{Chow:2013gba}. 

\paragraph{AdS$_2$ background.} The attractor solution for this example is very simple. The scalars simplify to
\be
\bar \chi_i=0 ~,\qquad \bar\varphi_i=0 ~.
\ee
Solving the attractor equations  for this background gives
\be\label{eq:attc6}
\begin{aligned}
 \frac{1}{\ell_2^2}+\frac{1}{\Phi^2_0}   = \frac{(Q_1)^2}{\Phi_0^4}~,\qquad
\frac{1}{\ell_2^2}-\frac{1}{\Phi^2_0}=6g^2  ~.
\end{aligned}
\ee
Note that these equations imply that $\Phi_0^2>\ell_2^2$ and $6g^2\ell_2^2<1$. The solution for the size of the $S^2$ is
\be
6g^2\Phi_0^2= -1 +\sqrt{1+6g^2 (Q_1)^2} ~.
\ee

\paragraph{Spectrum of operators.} The linearized equations \eqref{eq:linearphi} and \eqref{eq:linearchi} are also very simple. They become
\be\label{eq:bchsingleQ}
\begin{aligned}
\bar{\square} \bch_i - (4 g^2 + \frac{2}{\Phi_0^2}) \bch_i=0~,
\end{aligned}
\ee
and
\be\label{linearcase6}
\begin{aligned}
\bar{\square} \bph_1 - (4 g^2 + \frac{2}{\Phi_0^2}) \bph_1 + (6 g^2 + \frac{2}{\Phi_0^2}) \bph_2 + (6 g^2 + \frac{2}{\Phi_0^2}) \bph_3=0~,\\
\bar{\square} \bph_2 - (4 g^2 + \frac{2}{\Phi_0^2}) \bph_2 - (6 g^2 + \frac{2}{\Phi_0^2}) \bph_3 + (6 g^2 + \frac{2}{\Phi_0^2}) \bph_1=0~,\\
\bar{\square} \bph_3 - (4 g^2 + \frac{2}{\Phi_0^2}) \bph_3 - (6 g^2 + \frac{2}{\Phi_0^2}) \bph_2+ (6 g^2 + \frac{2}{\Phi_0^2}) \bph_1=0~.
\end{aligned}
\ee
 Note that in this case there are no terms proportional to $\Y$, and hence there are no inhomogeneous solutions ($\vec a=0$). 
 
The spectrum of conformal dimensions is the following. The masses for $\bch_i$ are easy to read off from \eqref{eq:bchsingleQ}, and hence we have three eigenstates $\mathfrak{Z}_i, i = 3,4,5$\footnote{The labelling of eigenstates is such that the conformal dimensions increase as we go from $\mathfrak{Z}_1$ to $\mathfrak{Z}_6$.} with conformal dimension
\be
\begin{aligned} \label{eq:Deltachi}
\Delta_{\chi}&= \frac{1}{2}+ \sqrt{ \frac{1}{4} + \ell_2^2m_\chi^2} \\
&=\frac{1}{2}+ \sqrt{ \frac{9}{4} -8 \ell_2^2g^2} ~.
\end{aligned}
\ee
For $0\leq6g^2\ell_2^2< 1$, we have $1.46<\Delta_{\chi}\leq2$ and hence the $\bch_i$ are less irrelevant than $\Y$: the effect of the AdS$_4$ ($g\neq0$) surrounding is to lower the conformal dimension relative to the ungauged case. Next, diagonalizing \eqref{linearcase6}, the three eigenvalues and eigenstates are 
    \begin{equation} \label{eq:singlepqmass}
    \begin{aligned}[c]
      &m_1^2  = -2g^2~, &&\Delta_1= {1\over 2}+\sqrt{\frac{1}{4}-2g^2\ell_2^2} ~: ~&&\quad
      \begin{array}{l} 
      {\mathfrak{Z}}_1 = \bph_1 + \bph_3 ~, \\
        {\mathfrak{Z}}_2 = \bph_1 + 2\bph_2 - \bph_3~,
      \end{array}\\[0.5em]
      &m_2^2 =16 g^2 + \frac{6}{\Phi_0^2}~, &&\Delta_2= {1\over 2}+\sqrt{\frac{25}{4}-20g^2\ell_2^2}~: ~&&\quad
       \begin{array}{l} 
      \displaystyle {\mathfrak{Z}}_6 = -\bph_1 + \bph_2 + \bph_3 ~.
      \end{array} 
    \end{aligned}
    \end{equation}
Again, conformal dimensions are lowered due to the presence of $g$. But more importantly, in this sector we have two negative mass-squared states, which could trigger an instability. 
Demanding that $m_1^2$ complies with the Breitenlohner-Freedman bound in AdS$_2$, requires that 
\be
8 g^2 \ell_2^2 \leq 1~,~~ \textrm{or equivalently,}~~~~\quad 2 g^2 \Phi_0^2 \leq 1~.
\ee
In terms of the electric charge this means
\be
g^2 Q_1^2 \leq  \frac{5}{2}~.
\ee
Hence only black holes with very small charges (relative to the AdS$_4$ radius) are stable. 

\paragraph{Interactions.} Finally, we report on the corrections to the two-point functions due to a non-trivial background value of $\Y$. The corrections are of the form \eqref{two:total}. We find 
    \begin{equation} \label{eq:singlepq}
    \begin{aligned}[c]
    \Delta_{\chi}~: ~ & \hat D_\chi = - \frac{3}{4(\Delta_\chi -1)} \le( -\frac{4}{\ell_2^2} + 16g^2\ri)~,
    \\[0.5em]
      \Delta_1 ~: ~  & \hat D_1= - \frac{3}{4(\Delta_1 -1)} \le(\frac{1}{\ell_2^2}+g^2\ri)~,
        \\[0.5em]
    \Delta_2~: ~  & \hat D_2= - \frac{3}{4(\Delta_2 -1)} \le(- \frac{14}{\ell_2^2}+46g^2\ri)~,
    \end{aligned}
    \end{equation}
where $\hat D$ is defined in \eqref{eq:Dhat}. Computing them explicitly using the conformal dimensions in \eqref{eq:Deltachi} and \eqref{eq:singlepqmass}, we find that both $\hat D_\chi$ and $\hat D_2$ are positive for $0 \leq g^2 < \frac{1}{6\ell_2^2}$:
    \begin{equation}
        \begin{aligned}
             2.19  < \ell_2^2 &\hat D_\chi \leq 3 ~, \\
             3.93  < \ell_2^2 &\hat D_2 \leq  \frac{21}{4}~.
        \end{aligned}
    \end{equation}
Again the effect of $g$ is to lower the values relative to the ungauged theory. Setting $g=0$ matches $\hat D_\chi$ with \eqref{eq:Dbps} and $\hat D_2$ with \eqref{eq:tD6mass}. The case $\hat D_1$ is similar to $\hat D_1$ in the magnetic non-BPS case with $P^1 = - P^4$, i.e.\ \eqref{eq:oppositeDtildes}: for $g = 0$, the correlator is extremal and $K_{\Y \mathfrak{Z}_4\mathfrak{Z}_4}$ diverges; for $g^2 > \frac{1}{8\ell_2^2}$ the BF bound is violated. In the range $g^2 \in (0, \frac{1}{8\ell_2^2})$, $\hat D_1$ is positive. 

\section{Discussion}\label{sec:disc}

{\renewcommand{\arraystretch}{1.6}
\begin{table}
\hspace*{-1.4cm}
\begin{tabular}{c|c|c|c|l|c|}
\cline{4-6}
\multicolumn{3}{l|}{}                                                                                                  & \multicolumn{2}{c|}{\bf Spectrum}       & \multicolumn{1}{c|}{\bf Interactions} \\ \cline{3-6} 
\multicolumn{2}{l|}{}                                                       & \multicolumn{1}{c|}{\bf $(Q_I, P^I)$}                                 & {\bf $\Delta$ } & \multicolumn{1}{c|}{\bf Eigenstates} & $\expval{\mathfrak{Z} \mathfrak{Z}} = \expval{\mathfrak{Z} \mathfrak{Z}}_{\rm free} (1 + \hat D (...) ) $ \\ \hline
\multicolumn{1}{|c|}{\multirow{7}{*}{\rotatebox[origin=c]{90}{\bf ungauged}}} & \rotatebox[origin=c]{90}{{~~} BPS { ~~}}                       &  \makecell{$Q_I \neq 0$, \\ $P_I \neq 0$}               & $\Delta_\mathfrak{Z} = 2$                     &   \multicolumn{1}{c|}{$\vec{\mathfrak{Z}} = (\bph_i, \bch_i)$}          &    $\hat D_\mathfrak{Z} = \frac{3}{\ell_2^2}  $                             \\ \cline{2-6} 
\multicolumn{1}{|c|}{}                          & \multirow{5}{*}{\rotatebox[origin=c]{90}{non-BPS}}  & \multirow{3}{*}{\makecell{\eqref{eq:nonBPSequalPQs1}, 7 in- \\ dependent \\ parameters}}        &  $\Delta_1 = 1$             &          $
     \left\{ \begin{array}{l} \mathfrak{Z}_1 = \bch_1 + \bch_3 \\ \mathfrak{Z}_2 = c_1 \bph_2 + \bch_1 + c_2 \bch_2 \end{array}\right.$      &  $\hat D_1$ undetermined                              \\
\multicolumn{1}{|c|}{}                          &                           &                      &  $\Delta_2 = 2$                      & $\left\{ \begin{array}{l} \mathfrak{Z}_3 = \bph_1 \\ \mathfrak{Z}_4 = \bph_3 \\ \mathfrak{Z}_5 = c_3 \bph_2 + c_4 \bch_2 \end{array} \right.$            &    $\hat D_2 =  \frac{3}{\ell_2^2}  $                 \\
\multicolumn{1}{|c|}{}                          &                           &                      &  $\Delta_3 = 3$                      & $\mathfrak{Z}_6 = c_1 \bph_2 - \bch_1 + c_2 \bch_2 + \bch_3$            &   $\hat D_3 = \frac{21}{4\ell_2^2}$           \\ \hline
\multicolumn{1}{|l|}{\multirow{18}{*}{\rotatebox[origin=c]{90}{\bf gauged}}}  & \multirow{5}{*}{\rotatebox[origin=c]{90}{BPS}}      &   \multirow{4}{*}{\makecell{Magnetic, \\ $\mathfrak{n}_1 = 2 - 3 \mathfrak{n}_4$ \\${\mathfrak{n}_2 = \mathfrak{n}_3 = \mathfrak{n}_4}$ } }                       & $\begin{array}{l} 0.35<\Delta_1^- < 0.42 ~, \\[-0.2cm] 0.58<\Delta_1^{+}<0.65 \end{array}$                     &  $\left\{ \begin{array}  {l} 
      {\mathfrak{Z}}_1 = -\bch_1 + \bch_3  \\
        {\mathfrak{Z}}_2 = -\bch_1 + 2\bch_2 - \bch_3 \end{array} \right. $        &    $\begin{array}{l}  10.3\,g^2 < \hat D_1^- < 12.4\,g^2~, \\[-0.2cm] 14.1\,g^2 < \hat D_1^{+} < 23.5\,g^2 \end{array}$ \\
\multicolumn{1}{|l|}{}                          &                           &           &    $1.15 < \Delta_2 < 1.31$        &    $\mathfrak{Z}_3 = \bph_1 + \bph_2 + \bph_3$         &    $-19.7\,g^2 < \hat D_2 <  -4.49\,g^2 $\\
\multicolumn{1}{|l|}{}                          &                            &           & $ 1.57 < \Delta_3 < 1.65$                     &   $\left\{ \begin{array}{l} {\mathfrak{Z}}_4 = -\bph_1 + \bph_3 \\
        {\mathfrak{Z}}_5 = -\bph_1 + 2\bph_2 - \bph_3 \end{array} \right.$          &   $25.3\,g^2 < \hat D_3 < 34.1\,g^2$                                \\
\multicolumn{1}{|l|}{}                          &                           &           & $2.15<\Delta_4 <2.31$                     &  $\mathfrak{Z}_6 = \bch_1 + \bch_2 + \bch_3$            &    $35.4\,g^2 < \hat D_4 < 48.7\,g^2 $                               \\ \cline{2-6} 
\multicolumn{1}{|l|}{}                          & \multirow{10}{*}{\rotatebox[origin=c]{90}{non-BPS}} & \makecell{ Magnetic, \\ $P^1 = P^2 = P^3 $ \\ $P^4 = P^1$}    & $1.46 < \Delta_\mathfrak{Z} \leq 2 $                     &  \multicolumn{1}{c|}{$\vec{\mathfrak{Z}} = (\bph_i, \bch_i)$}          &       $2.19 < \ell_2^2 \hat D_\mathfrak{Z}  \leq 3$                            \\ \cline{3-6} 
\multicolumn{1}{|l|}{}                          &                           & \multirow{3}{*}{\makecell{Magnetic, \\ $P^1 = P^2 = P^3 $\\$P^4 = - P^1$}} &  $\Delta_1^* \leq 1$                     &   $\left\{ \begin{array}  {l} 
      {\mathfrak{Z}}_1 = -\bch_1 + \bch_3  \\
        {\mathfrak{Z}}_2 = -\bch_1 + 2\bch_2 - \bch_3 \end{array} \right.$          &    $\hat D_1 > 0$ \\
\multicolumn{1}{|l|}{}                          &                           &                                          & $1.46 <\Delta_\varphi \leq 2 $                    &   $\mathfrak{Z}_{i+2} = \bph_i, ~ i = 1,2,3$            &      $2.19 < \ell_2^2 \hat D_{\varphi}  \leq 3$                               \\
\multicolumn{1}{|l|}{}                          &                           &                                         & $ 2.2 < \Delta_4 \leq 3$                     &   $\mathfrak{Z}_6 = \bch_1 + \bch_2 + \bch_3$          &    $3.93 < \ell_2^2 \hat D_4 \leq \frac{21}{4}$                               \\ \cline{3-6} 
\multicolumn{1}{|l|}{}                          &                           &   \multirow{3}{*}{{\makecell{Dyonic, \\ $P^1 = \pm Q_1$ \\ $ P^{I\neq 1}= 0$\\ $Q_{I \neq 1} =0$}}}                & $\Delta_1^* \leq 1 $                     & $\left\{ \begin{array}{l} \mathfrak{Z}_4 = \bph_1 + \bph_3 \\ \mathfrak{Z}_5 = \bph_1 + 2\bph_2 - \bph_3 \end{array} \right.$            &    $\hat D_1 > 0$                               \\
\multicolumn{1}{|l|}{}                          &                           &                &  $1.46 <\Delta_\chi \leq 2$                     &  $\mathfrak{Z}_i = \bch_i, ~{i = 1,2,3}$             &   $2.19 < \ell_2^2 \hat D_{\chi}  \leq 3$                                \\
\multicolumn{1}{|l|}{}                          &                           &                                          & $2.2< \Delta_2 \leq 3$                     &   $\mathfrak{Z}_6 = -\bph_1 + \bph_2 + \bph_3$          &    $3.93 < \ell_2^2 \hat D_2 \leq \frac{21}{4}$                               \\ \hline
\end{tabular}
\captionsetup{width=\textwidth}
\caption{A summary of our results for the different cases we considered: ungauged/gauged, BPS/non-BPS, and charge configurations. We report (the range of) the conformal dimensions, the eigenstates, and the corrections to the two-point functions as defined in \eqref{two:total} with $\hat D$ given in \eqref{eq:Dhat}. The ranges are for $0 \leq 6 g^2 \ell_2^2 < 1$. The symbol $\Delta^*$ indicates that the corresponding states can violate the BF bound if $8 g^2 \ell_2^2 <1$, and that $\Delta = 1$ for $g = 0$; in those cases, we only give the sign of $\hat D$. Note that the $c_i$ in the ungauged, non-BPS eigenstates, can be read off from \eqref{eq:nonBPSEV}. We dropped the superscript $({\rm hom})$ for the magnetic gauged cases, and presented the orthogonal (but not orthonormal) eigenstates, for clarity and brevity. }\label{tab:summary}
\end{table}
}

We explored aspects of near-AdS$_2$ backgrounds that appear in ${\cal N}=2$, $D=4$ supergravity. We focused on quantifying the spectrum of operators, and their interactions with the JT sector. In table \ref{tab:summary} we collected the most prominent examples studied here, and summarized their main features. From this table  there are a few important lessons:
\begin{itemize}
    \item {\it Supersymmetry is key.} In every single non-BPS example we considered there is something undesirable: either we have unstable modes in the gauged theory, or we have problems with the extremal three-point correlators in the ungauged case. BPS backgrounds have a well-defined EFT description in all cases. 
    \item {\it Gauged versus ungauged.} Not surprisingly, the effect of AdS$_4$, relative to Minkowski$_4$, on the spectrum of AdS$_2$ operators is to lower the conformal dimensions.\footnote{This also happens in AdS$_5$; see e.g. \cite{Castro:2018ffi}.} This  allows for the presence of operators that are relevant, or even unstable, for the gauged theory. It is one reflection of how the surrounding (UV embedding) has an imprint on the IR physics.  
    \item {\it Who is relevant, marginal and irrelevant.} It is interesting to see how the spectrum for a black hole can be plain and simple (BPS ungauged theory), or have all flavours of operators available (BPS gauged theory). This is an indication that the ingredients that go into building a statistical description of the black hole will not be universal. 
    \item {\it Expected and unexpected pathologies.} One expected pathology we encountered in our analysis is the presence of modes that violate the BF bound for backgrounds in the gauged theory. This is a common occurrence in AdS$_2\times \mathbb{R}^2$ in the context of AdS/CMT { \cite{Denef:2009tp,Iqbal:2011ae,Hartnoll:2016apf}}, although less discussed for AdS$_2\times S^2$ \cite{Gubser:2008px,Sonner:2009fk}. 
    The unexpected pathology is the non-vanishing extremal cubic coupling among the marginal operator and $\Y$ for non-BPS backgrounds in the ungauged theory, as discussed in Sec.\,\ref{sec:ungaugedInt}. Although it is well known that non-BPS black holes have a flat direction in the attractor mechanism, it is disappointing that this spoils the construction of an effective field theory around near-AdS$_2$.
\end{itemize}
Of course here we focused on specific holographic aspects of the near-AdS$_2$ backgrounds: the spectrum, a specific type of interaction, and the imprint on one specific correlation function. Other interactions, other entries in the holographic dictionary, and how they affect other observables would be interesting to study. A few possible future directions are the following.
 
\paragraph{From UV to IR.} 
One of the most interesting, and challenging, directions to pursue is to reproduce the results of table \ref{tab:summary} from a dual description. This might be a feasible task for supersymmetric black holes, where we have some control already on their extremal entropy, such as the BPS black holes in the gauged and ungauged theory. For the gauged cases, it would be very interesting if one could reproduce the values of $\Delta$ from the dual CFT$_3$ in the UV.  
 
In addition to the spectrum, we also reported on the cubic interactions appearing in the near-AdS$_2$ region and their imprint on the two-point functions. This is a non-universal entry in the holographic dictionary, and hence contains valuable information about the near-CFT$_1$. One direction to pursue is to connect $\hat D$ to an observable outside the near-horizon region, and hence tie its value and properties to a correction that can be computed in the UV description of the black hole. This should be conceptually clear, although technically cumbersome to evaluate; if feasible, it would provide a non-trivial check of our analysis. 
 
\paragraph{Imprint on quantum and higher derivative corrections.} A natural question is how the operator content we quantified affects the quantum entropy of black holes. In particular, we found in the gauged theory that there are matter fields with $\Delta<3/2$, which according to \cite{Maldacena:2016hyu} are the dominant effect over the Schwarzian sector. It would be interesting to understand the connection to \cite{Heydeman:2020hhw,Iliesiu:2020qvm}, and have a more refined understanding of the statistical system. 

It is also important to contrast thermodynamic properties of BPS versus non-BPS black holes beyond the area law. Several issues already arose in our analysis, and it will be useful to place those in the context of quantum corrections to the black hole entropy. For example, for non-extremal black holes the logarithmic corrections behave very differently, as illustrated in \cite{Charles:2015eha} versus \cite{CastroGodetLarsenEtAl2018}. Also the results of \cite{Heydeman:2020hhw,Iliesiu:2020qvm} show important differences between BPS and non-BPS states, which quantifies the role of the mass gap in both cases. 

Finally, it would be interesting to quantify the role of higher derivative corrections in ${\cal N}=2$ supergravity in our analysis.  This was recently revisited for AdS$_2$ backgrounds for the ungauged theory in \cite{Aniceto:2020saj}, and there are as well some new developments in the gauged theory done by \cite{Bobev:2021oku,Genolini:2021urf}.

\paragraph{Black hole zoo.} As stressed from the beginning, we only focused on backgrounds that are a direct product of AdS$_2$ and $S^2$; and within the gauged theory, we considered a small subset of all solutions. It would clearly be interesting to expand this analysis and establish if the patterns found here are robust for all dyonic solutions in AdS$_4$. 

Of course there is also a much richer space of solutions that deserve attention. For instance, the inclusion of different horizon topologies, which is quite intricate in the gauged theory; several AdS$_2$ backgrounds of this form have been recently discussed in \cite{Ferrero:2021ovq, Couzens:2021rlk, Suh:2021hef}, including an overview of possible solutions in the 4D gauged theory. Rotation is also a very interesting aspect to explore, and our current work was inspired by the new features found in five-dimensional rotating black holes \cite{Castro:2018ffi,Castro:2021fhc}. In contrast to the results in \cite{Castro:2021fhc}, we did not see a change of sign for $\tilde D$ for a fixed operator, which might be attributed to the lack of rotation here. Understanding and decoding the patterns in  $\tilde D$ clearly deserves more attention. In four dimensions rotation is a much more difficult parameter to introduce in near-AdS$_2$, as reflected by the analysis of near-extremal Kerr in \cite{Castro:2021csm}. Still it would be interesting to understand how the interplay between rotation and supersymmetry enters in our discussion, and in particular to connect it with the developments in \cite{Anninos:2017cnw,Larsen:2020lhg,David:2020jhp}.

\paragraph{Integrability conditions on non-extremal black holes.}
As shown in \cite{Chow:2013gba}, which builds upon observations in \cite{Lu:2013ura}, a generic non-extremal dyonic black hole background is not physically reasonable for  $g\neq0$ and arbitrary values of $({Q}_I, {P}^I)$. In a nutshell, non-extremal AdS$_4$ black holes cannot carry both magnetic and electric charge without imposing a constraint. One way to discover this inconsistency comes from demanding integrability of the mass of the black hole. Integrable conserved charges are paramount for a well-defined phase space and the validity of the first law of thermodynamics. It would be interesting to understand how this integrability condition is present in the context of near-AdS$_2$. The AdS$_2$ background will exist for any values of $({Q}_I, {P}^I)$, but there should be a restriction on how to deform away from it such that this deformation leads to an integrable non-extremal solution.

\section*{Acknowledgements}
We are grateful to Alex Belin, Jan de Boer, Geoffrey Comp\`ere, Roberto Emparan, Nabil Iqbal, and Chiara Toldo for discussions on this topic.  The work of AC and EV is supported in part by the Delta ITP consortium, a program of the NWO that is funded by the Dutch Ministry of Education, Culture and Science (OCW). The work of EV is part of the research programme of the Foundation for Fundamental Research on Matter (FOM), which is financially supported by NWO.

\appendix

\section{Conventions}\label{app:conventions}

 Here we gather some basic conventions used in Sec.\,\ref{sec:gaugedsugra} for easy comparison with other references. For a $p$-form $\omega$ with components defined by 
	\begin{equation}
		\omega = \frac{1}{p!} \omega_{\mu_1 \cdots \mu_p} \dd x^{\mu_1} \wedge \cdots \wedge \dd x^{\mu_p}~, 
	\end{equation}
the dual $\star \omega$ is 
	\begin{equation}
		\star \omega = (\star \omega )_{\nu_1 \cdots \nu_{n - p}} \dd x^{\nu_1} \wedge \cdots \wedge \dd x^{\nu_{n-p}}~,
	\end{equation}
where
	\begin{equation}
		(\star \omega )_{\nu_1 \cdots \nu_{n-p}} = \frac{1}{p!} \epsilon_{\nu_1 \cdots \nu_{n-p} \mu_1 \cdots \mu_p} \omega^{\mu_1 \cdots \mu_p}~.
	\end{equation}
Here $n$ is the number of spacetime dimensions. 

The four-dimensional Levi-Civita tensor  is given by
	\begin{equation}
	\epsilon_{\mu\nu\alpha\beta} = \sqrt{-g^{(4)}} \tilde{\epsilon}_{\mu\nu\alpha\beta}~, 
	\end{equation}
where $\tilde{\epsilon}$ is the Levi-Civita symbol $\tilde{\epsilon}_{0123} = 1$. Also, it is useful to recall that in four-dimensions, in Lorentzian signature, the top form is
\begin{align}
\dd x^\alpha \wedge\dd x^\beta \wedge \dd x^\mu \wedge \dd x^\nu &= - \epsilon^{\alpha\beta\mu\nu} \sqrt{-g^{(4)}}\dd x^1 \wedge\dd x^2 \wedge \dd x^3 \wedge \dd x^4\cr
&= -\epsilon^{\alpha\beta\mu\nu} \sqrt{-g^{(4)}} d^4x~.
\end{align}
The two-dimensional Levi-Civita tensor  is given by
	\begin{equation}
	\epsilon_{ab} = \sqrt{-g^{(2)}} \tilde{\epsilon}_{ab}~, 
	\end{equation}
where $\tilde{\epsilon}$ is the Levi-Civita symbol $\tilde{\epsilon}_{01} = 1$. 

\section{Aspects of \texorpdfstring{$U(1)^4$}{U14} supergravity}\label{app:Fterms}
In this appendix we collect various formulas that are used that are specific to $U(1)^4$ supergravity. In particular we present explicit formulas related to integrating out the fields strengths in our setup. We also present the explicit attractor solutions for this theory.

We start by writing out explicitly  the matrices introduced in \eqref{eq:maxwell-action-mod}. Writing $h_{IJ}$ and $k_{IJ}$ in matrix notation, we have
  \be\label{eq:hmatrix}
H \equiv  -\frac{\chi_1}{2} 
\begin{pmatrix}
  0&0&0&1\\
  0&0&1 &0\\
  0&1&0&0\\
  1&0&0&0
  \end{pmatrix}~,
  \ee
 and
   \be
K\equiv 
\begin{pmatrix}
  k_{11}&k_{12}&k_{13}& k_{14}\\
  k_{12}&k_{22}&k_{23}& k_{24}\\
  k_{13}&k_{23}&k_{33}& k_{34}\\
  k_{14}&k_{24}&k_{34}& k_{44}\\
  \end{pmatrix}~,
  \ee
  with
  \be\label{eq:kmatrix}
  \begin{aligned}
  k_{11}&= e^{-\varphi_1+\varphi_2+\varphi_3}~,\\
  k_{12}&= e^{-\varphi_1+\varphi_2+\varphi_3} \chi_3~,\\
  k_{13}&= e^{-\varphi_1+\varphi_2+\varphi_3}\chi_2~,\\
  k_{14}&=-e^{-\varphi_1+\varphi_2+\varphi_3}\chi_2\chi_3~,\\
  k_{22}&=e^{-\varphi_1+\varphi_2+\varphi_3} \chi_3^2 +e^{-\varphi_1+\varphi_2-\varphi_3}~, \\
  k_{23}&= e^{-\varphi_1+\varphi_2+\varphi_3}\chi_2\chi_3 ~,\\
  k_{24}&= -e^{-\varphi_1+\varphi_2+\varphi_3}\chi_2\chi_3^2 -e^{-\varphi_1+\varphi_2-\varphi_3}\chi_2~,\\
  k_{33}&=e^{-\varphi_1+\varphi_2+\varphi_3} \chi_2^2 +e^{-\varphi_1-\varphi_2+\varphi_3} ~,\\
  k_{34}&= -e^{-\varphi_1+\varphi_2+\varphi_3}\chi_2^2\chi_3 -e^{-\varphi_1 - \varphi_2+\varphi_3}\chi_3~,\\
  k_{44}&= e^{-\varphi_1+\varphi_2+\varphi_3} \chi_2^2\chi_3^2 +e^{-\varphi_1+\varphi_2-\varphi_3}\chi_2^2+e^{-\varphi_1-\varphi_2+\varphi_3}\chi_3^2+e^{-\varphi_1-\varphi_2-\varphi_3}~.\\
  \end{aligned}
  \ee
A few identities that these matrices satisfy are
\be\label{eq:idhk1}
H^2= \frac{\chi_1^2}{4} \mathds{1}_{4\times4}~,
\ee
i.e.\ $H$ is proportional to its own inverse, and
\be\label{eq:idhk2}
e^{-\varphi_1} H\, K^{-1}= e^{\varphi_1}K\, H~.
\ee
Using these identities we can write the potential introduced in \eqref{eq:defV} as
\be\label{potential}
V({\bf P},{\bf Q})\equiv ({\bf P}^I ~~  {\bf Q}_I)
   \begin{pmatrix}
    (1+\chi_1^2 e^{2\varphi_1})k_{IJ} &- 2 e^{2\varphi_1}(kh)_I^{~\,J}\\
  - 2 e^{2\varphi_1} (hk)^{I}_{~J} & (k^{-1})^{IJ}\\
 \end{pmatrix}
   \begin{pmatrix}
 {\bf P}^J  \\
 {\bf Q}_J
 \end{pmatrix} ~.
\ee
When manipulating the linearized equations of motion for the scalars, it will be also useful to note that for the $U(1)^4$ theory the potential also obeys
	\begin{equation}\label{eq:idV}
		\partial_{\varphi_i}^2 V = V~, \quad \partial_{\varphi_i}\partial_{\chi_i} V = \partial_{\chi_i} V~, \quad \partial_{\chi_1}^2 V = 2 e^{2\varphi_1} {\bf P}^I k_{IJ} {\bf P}^J~.
	\end{equation}
When solving for the equations of motion \eqref{eq:eomF}, the $x^a$ components of the modified field strengths  explicitly are
	\be\label{eq:solcF}
	\begin{aligned}
		{\cal F}^1_{~ab} &= \Phi^{-3} (Q_1 + \chi_1 P^4) e^{\varphi_1-\varphi_2-\varphi_3} \epsilon_{ab} \\
		\tilde{{\cal F}}_{2\, ab} &= \Phi^{-3} \left( P^2 - \chi_1 Q_3 - \chi_3Q_1 - \chi_3\chi_1 P^4\right) e^{\varphi_1-\varphi_2+\varphi_3} \epsilon_{ab} \\ 
		\tilde{{\cal F}}_{3\,ab} &= \Phi^{-3} \left( P^3 - \chi_1 Q_2 - \chi_2 Q_1 - \chi_2 \chi_1 P^4\right) e^{\varphi_1+\varphi_2-\varphi_3} \epsilon_{ab}  \\
		{\cal F}^4_{~ab} &= \Phi^{-3} \big( Q_4 - \chi_1 \chi_2 Q_3- \chi_1  \chi_3 Q_2  - \chi_2 \chi_3 Q_1 \cr 
		& \qquad \qquad - \chi_1\chi_2\chi_3 P^4 + \chi_3 P^3 + \chi_2 P^2 + \chi_1 P^1 \big)  e^{\varphi_1+\varphi_2+\varphi_3} \epsilon_{ab}~,
	\end{aligned}
	\ee
and we can transform them to the usual field strengths $F^I = \dd A^I$ and $\tilde{F}_I = \dd \tilde{A}_I$ via \eqref{eq:dicFcF}. The components along the 2-sphere are given explicitly in \eqref{eq:finalF}. 

\subsection{\texorpdfstring{AdS$_2$}{AdS2} backgrounds}\label{app:ads2}

In this section we write more explicitly the equations that determine the AdS$_2$ backgrounds of Sec.\,\ref{sec:IR}. To start, it is useful to introduce some more notation. 
We will define for the magnetic charges
	\begin{equation} \label{eq:calPs}
	\begin{aligned}
	{\cal P}^1\equiv & \, e^{\barp_2 + \barp_3} \left( P^1 - \barc_2 Q_3 - \barc_3 Q_2 - \barc_2 \barc_3 P^4 \right) ~,\\
	{\cal P}^2\equiv & \, e^{\barp_1 + \barp_3} \left(P^2 - \barc_1 Q_3 - \barc_3Q_1 - \barc_3\barc_1 P^4 \right)~,\\
	{\cal P}^3\equiv  & \,e^{\barp_1 + \barp_2} \left(P^3 - \barc_1 Q_2 - \barc_2 Q_1 - \barc_2 \barc_1 P^4 \right)~,\\
	{\cal P}^4\equiv & \,P_4~,
	\end{aligned}
	\end{equation}
and for the electric charges
for $Q_{1,2,3}$ we have
\be \label{eq:calQsA}
\begin{aligned}
{\cal Q}_i\equiv e^{\barp_i} \left( Q_i + \barc_i P^4 \right)~,\\
\end{aligned}
\ee
while for the fourth charge 
\be \label{eq:calQsB}
{\cal Q}_4\equiv e^{\barp_1 + \barp_2 + \barp_3} \left( Q_4 - \barc_1 \barc_2 Q_3- \barc_1  \barc_3 Q_2  - \barc_2 \barc_3 Q_1 - \barc_1\barc_2\barc_3 P^4 + \barc_3 P^3 + \barc_2 P^2 + \barc_1 P^1 \right)~.
\ee
These are shifts that are completing the squares in $V({\bf P},{\bf Q})$, i.e., making the matrix in \eqref{potential} diagonal; and they also follow from the definitions of ${\cal F}^I$ and ${\cal F}_I$ in \eqref{eq:solcF}.

Using ${\cal Q}_I$ and ${\cal P}^I$, the equations for $\varphi_i$ in \eqref{att:varphi} reduces to
	\begin{equation} \label{eq:atteqsscalars}
	\begin{aligned}
		({\cal P}^4)^2 - ({\cal P}^3)^2 - ({\cal P}^2)^2 + ({\cal P}^1)^2 - {\cal Q}_1^2 + {\cal Q}_2^2 + {\cal Q}_3^2 - {\cal Q}_4^2 + 2 e^{\barp_1+\barp_2+\barp_3} g^2 \Phi_0^4(2\sinh \barp_1 + \barc_1^2 e^{\barp_1}) &=0~, \\
		({\cal P}^4)^2 - ({\cal P}^3)^2 + ({\cal P}^2)^2 - ({\cal P}^1)^2 + {\cal Q}_1^2 - {\cal Q}_2^2 + {\cal Q}_3^2 - {\cal Q}_4^2 + 2 e^{\barp_1+\barp_2+\barp_3} g^2 \Phi_0^4(2\sinh \barp_2 + \barc_2^2 e^{\barp_2})&=0 ~, \\
	 	({\cal P}^4)^2 + ({\cal P}^3)^2 - ({\cal P}^2)^2 - ({\cal  P}^1)^2 + {\cal Q}_1^2 + {\cal Q}_2^2 - {\cal Q}_3^2 - {\cal Q}_4^2 + 2 e^{\barp_1+\barp_2+\barp_3} g^2 \Phi_0^4 (2\sinh \barp_3 + \barc_3^2 e^{\barp_3}) &=0~.
	\end{aligned}
	\end{equation}
The equations for the axions \eqref{att:chi} reduce to the following 
\begin{equation} \label{eq:atteqsaxions}
\begin{aligned}
		- {\cal P}^4 {\cal Q}_1 + {\cal P}^3 {\cal Q}_2 + {\cal P}^2 {\cal Q}_3 - {\cal P}^1 {\cal Q}_4 + 2 e^{\barp_1 +\barp_2+\barp_3} g^2 \Phi_0^4 \barc_1 = 0~, \\
		- {\cal P}^4 {\cal Q}_2 + {\cal P}^3 {\cal Q}_1 + {\cal P}^1 {\cal Q}_3 - {\cal P}^2 {\cal Q}_4 + 2 e^{\barp_1 +\barp_2+\barp_3} g^2 \Phi_0^4 \barc_2 = 0~, \\	
		- {\cal P}^4 {\cal Q}_3 + {\cal P}^2 {\cal Q}_1 + {\cal P}^1 {\cal Q}_2 - {\cal P}^3 {\cal Q}_4 + 2 e^{\barp_1 +\barp_2+\barp_3} g^2 \Phi_0^4 \barc_3 = 0~,
\end{aligned}
\end{equation}
The equations in \eqref{eq:adsphi} read in this notation
    \begin{equation}\label{eq:L2equation}
    \begin{aligned}
        \frac{1}{\ell_2^2} &= \frac{1}{4\Phi_0^4}e^{-\barp_1 - \barp_2 - \barp_3} \Big( {\cal Q}_1^2 + {\cal Q}_2^2 + {\cal Q}_3^2 + {\cal Q}_4^2 + ({\cal P}^4)^2 + ({\cal P}^3)^2 + ({\cal P}^2)^2 +  ({\cal P}^1)^2 \Big) \\
        &\quad + \frac{g^2}{2} \sum_i (2\cosh \barp_i + \barc_i^2 e^{\barp_i} )~, 
    \end{aligned}
    \end{equation}
and
    \begin{equation} \label{eq:Phi0equation}
    \begin{aligned}
        \frac{1}{\Phi_0^2} &= \frac{1}{4\Phi_0^4}e^{-\barp_1 - \barp_2 - \barp_3} \Big( {\cal Q}_1^2 + {\cal Q}_2^2 + {\cal Q}_3^2 + {\cal Q}_4^2 + ({\cal P}^4)^2 + ({\cal P}^3)^2 + ({\cal P}^2)^2 +  ({\cal P}^1)^2 \Big) \\
        &\quad - \frac{g^2}{2} \sum_i (2\cosh \barp_i + \barc_i^2 e^{\barp_i} )~.
    \end{aligned}
    \end{equation}
A useful linear combination is:
	\begin{equation} \label{eq:l2combination}
		\frac{1}{\ell_2^2} - \frac{1}{\Phi_0^2} = g^2 \sum_i (2\cosh \barp_i + \barc_i^2 e^{\barp_i} )~.
	\end{equation}

\paragraph{Magnetic solution, non-BPS.} 
In the purely magnetic case, i.e.\ $Q_I = 0$ and $P^I \neq 0$, the attractor solution is very simple to write explicitly. All of the axions vanish at the horizon, 
    \begin{equation}
        \barc_i = 0~, 
    \end{equation}
and we can solve \eqref{eq:atteqsscalars} and \eqref{eq:Phi0equation} for the charges: 
\be\label{eq:magneticAttractor}
\begin{aligned}
({\cal P}^1)^2 &= {\Phi_0^4}\le( \frac{1}{\Phi^2_0}  +  g^2 (e^{- \bar \varphi_1} +e^{ \bar \varphi_2}+e^{\bar \varphi_3}) \ri) e^{\bar\varphi_1+\bar\varphi_2+\bar\varphi_3}~,\\
({\cal P}^2)^2 &= {\Phi_0^4}\le(\frac{1}{\Phi^2_0}  +  g^2(e^{ \bar \varphi_1} +e^{-  \bar \varphi_2}+e^{ \bar \varphi_3}) \ri) e^{\bar\varphi_1+\bar\varphi_2+\bar\varphi_3}~,\\
({\cal P}^3)^2&= {\Phi_0^4}\le(\frac{1}{\Phi^2_0} +  g^2 (e^{\bar \varphi_1} +e^{  \bar \varphi_2}+e^{- \bar \varphi_3})\ri) e^{\bar\varphi_1+\bar\varphi_2+\bar\varphi_3}~,\\
({\cal P}^4)^2 &= {\Phi_0^4}\le(\frac{1}{\Phi^2_0} + g^2(e^{- \bar \varphi_1} +e^{-\bar \varphi_2}+e^{-\bar \varphi_3})\ri) e^{\bar\varphi_1+\bar\varphi_2+\bar\varphi_3}~,
\end{aligned}
\ee
The attractor value of the dilaton $\Phi_0$ is given by
\be\label{eq:l2Magnetic}
 \frac{1}{\ell_2^2}  = \frac{1}{\Phi^2_0}+ 2g^2 \sum_i \cosh \bar \varphi_i ~.
\ee

\paragraph{Magnetic solution, BPS.} 
In \ref{sec:magneticBPS}, we discussed the BPS solution of \cite{Benini:2015eyy}, see also \cite{Cacciatori:2009iz,Hristov:2010ri}. The BPS equations in the notation of \cite{Benini:2015eyy} are\footnote{In these equations, we restored the gauge coupling $g^2$. Note that the coupling in \cite{Benini:2015eyy} differs from the one used here by a factor $\sqrt{2}$.} 
\be\label{eq:adsbps}
\begin{aligned}
0&=  g^2\le( L_1 + L_2 + L_3 +{1\over L_1 L_2 L_3}\ri) + {1\over \Phi_0^2} \le({\mathfrak{n}_1\over L_1} +{\mathfrak{n}_2\over L_2} +{\mathfrak{n}_3 \over L_3} + \mathfrak{n}_4 L_1L_2L_3\ri)\\
0&= g^2\le( L_1 + L_2 - L_3 -{1\over L_1 L_2 L_3}\ri) - {1\over \Phi_0^2} \le({\mathfrak{n}_1\over L_1} +{\mathfrak{n}_2\over L_2} -{\mathfrak{n}_3 \over L_3} - \mathfrak{n}_4 L_1L_2L_3\ri)\\
0&= g^2\le( L_1 - L_2 + L_3 -{1\over L_1 L_2 L_3}\ri) - {1\over \Phi_0^2} \le({\mathfrak{n}_1\over L_1} -{\mathfrak{n}_2\over L_2} +{\mathfrak{n}_3 \over L_3} - \mathfrak{n}_4 L_1L_2L_3\ri)\\
0&= g^2\le( L_1 - L_2 - L_3 +{1\over L_1 L_2 L_3}\ri) - {1\over \Phi_0^2} \le({\mathfrak{n}_1\over L_1} - {\mathfrak{n}_2\over L_2} - {\mathfrak{n}_3 \over L_3} + \mathfrak{n}_4 L_1L_2L_3\ri)\\
\frac{4g}{\ell_2}&=  g^2\le( L_1 + L_2 + L_3 +{1\over L_1 L_2 L_3}\ri) - {1\over \Phi_0^2} \le({\mathfrak{n}_1\over L_1} +{\mathfrak{n}_2\over L_2} +{\mathfrak{n}_3 \over L_3} + \mathfrak{n}_4 L_1L_2L_3\ri)
\end{aligned}
\ee
with the additional condition that 
\be\label{eq:condsum}
\sum_I \mathfrak{n}_I =2
\ee
and three of the $\mathfrak{n}_I$ are negative. Translating to our notation, which is done by comparing (A.2) there with \eqref{eq:L4D} here, we have 
\be \label{eq:ntoP}
\mathfrak{n}_1= g P^4~, \quad \mathfrak{n}_2 = g P^2~, \quad \mathfrak{n}_3 = g P^3~, \quad \mathfrak{n}_4 = g P^1~,
\ee
and 
\be \label{eq:Ltophis}
L_1 = e^{{1\over 2} (\varphi_1 +\varphi_2 +\varphi_3)}~,\quad L_2 = e^{{1\over 2} (-\varphi_1 +\varphi_2 -\varphi_3)}~,\quad L_3 = e^{{1\over 2} (-\varphi_1 -\varphi_2 +\varphi_3)}~.
\ee
Notice that \eqref{eq:adsbps} are linear in the charges, while the equations of motion \eqref{eq:atteqsscalars}-\eqref{eq:atteqsaxions} are quadratic. This is expected since the BPS equations are linear conditions, while an equation of motion is non-linear. This also reflects that BPS solutions are only a subset of the attractor solutions.

It is rather easy to solve \eqref{eq:adsbps} in a similar fashion as \eqref{eq:magneticAttractor}: this gives
\be\label{eq:cbps2}
\begin{aligned}
\mathfrak{n}_1&={g^{2} \Phi_0^2\over 2}\le(-{1\over L_2L_3} + L_1^2 -L_1L_2 - L_1L_3\ri)\\
\mathfrak{n}_2&={g^{2} \Phi_0^2\over 2}\le(-{1\over L_1L_3} + L_2^2 -L_2L_3 - L_2L_1\ri)\\
\mathfrak{n}_3&={g^{2} \Phi_0^2\over 2}\le(-{1\over L_1L_2} + L_3^2 -L_3L_2 - L_3L_1\ri)\\
\mathfrak{n}_4&={g^{2} \Phi_0^2\over 2}\le({1\over L_1^2 L_2^2L_3^2} -{1\over L_2L_3} -{1\over L_1L_2} -{1\over L_1L_3} \ri)\\
\end{aligned}
\ee
and 
    \begin{equation} \label{eq:L2Benini}
        \frac{1}{\ell_2^2} = \frac{g^2}{4} \left( \frac{1 + L_1^2 L_2 L_3 + L_1 L_2^3 + L_3 + L_1 L_2 L_3^2 }{L_1 L_2 L_3} \right)^2~.
    \end{equation}
A clever linear combination of \eqref{eq:cbps2} also gives $\Phi_0$ as
    \begin{equation} \label{eq:Phi0Benini}
        \Phi_0^2 = \frac{1}{2g^2}\left( \mathfrak{n}_4 L_1^2 L_2^2 L_3^2 - \mathfrak{n}_1 L_2 L_3 - \mathfrak{n}_2 L_1 L_3 - \mathfrak{n}_3 L_1 L_2 \right) ~.
    \end{equation}
The solutions \eqref{eq:cbps2} solve \eqref{eq:atteqsscalars}, and \eqref{eq:L2Benini} is just \eqref{eq:L2equation}. To satisfy \eqref{eq:Phi0equation} one must further impose \eqref{eq:condsum}. The condition on the negativity of three of the $\mathfrak{n}_I$ is an extra constraint that follows from  supersymmetry, i.e.\ the existence of a Killing spinor.
From \eqref{eq:cbps2} it is clear that the BPS solution given here is at a very different footing as compared to \eqref{eq:magneticAttractor}: all charges are proportional to $g$, so they are not smoothly connected to solutions in the ungauged theory. 

\section{Near-\texorpdfstring{AdS$_2$}{AdS2} properties of magnetic backgrounds}\label{app:magnetic}

In this appendix we collect results regarding the magnetic backgrounds studied in Sec.\,\ref{sec:gauged-cases}. The results in this section apply for both non-BPS (Sec.\,\ref{sec:magneticnonBPS}) and BPS (Sec.\,\ref{sec:magneticBPS}) cases. Here, we collect aspects of the linearized equations and interactions; the attractor solutions are described in App.\,\ref{app:ads2}. The expressions are given in the non-BPS notation; it is straightforward to translate them to the BPS language using the appropriate changes, notably \eqref{eq:ntoP}. 

For the purely magnetic case $Q_I = 0$ and $P^I \neq 0$,
the linearized equations in \eqref{eq:linearphi} become
\be
\begin{aligned}
&\bar{\square} \bph_i - 2(g^2\cosh\barp_j+g^2\cosh\barp_k + \frac{1}{\Phi_0^2}) \bph_i  \\
&\qquad + 2g^2 \sinh\barp_k \,\bph_j + 2g^2 \sinh\barp_j\,\bph_k+ \frac{8g^2}{\Phi_0}  \sinh\barp_i \Y=0~,
\end{aligned}
\ee
where $i\neq j\neq k$. We can split the solutions into homogeneous and inhomogeneous parts as in \eqref{eq:hominh}; the inhomogeneous parts of $\bch_i$ can consistently be set to vanish. Solving for $a_{\varphi,i}$ gives 
\be \label{eq:inhomMagnetic}
\begin{aligned}
\bph_i  = \frac{2}{\Phi_0} \frac{1 - e^{\barp_i + \barp_j} - e^{\barp_i + \barp_k}+ e^{\barp_j + \barp_k}}{1 + e^{\barp_1 + \barp_2} + e^{\barp_1 + \barp_3}+ e^{\barp_2 + \barp_3}}\Y + \bph_i^{\rm hom}~.
\end{aligned}
\ee
The homogeneous solutions $(\bph_i^{\rm hom}, \bch_i^{\rm hom})$ satisfy 
\be
\begin{aligned}
&\bar{\square} \bph_i^{\rm hom}- 2(g^2\cosh\barp_j+g^2\cosh\barp_k + \frac{1}{\Phi_0^2}) \bph_i^{\rm hom} \\
&\qquad + 2g^2 \sinh\barp_k \,\bph_j^{\rm hom} +  2g^2 \sinh\barp_j\,\bph_k^{\rm hom}=0~,
\end{aligned}
\ee
and
\be
\begin{aligned}
&\bar{\square} \bch_i^{\rm hom} - 2(g^2\cosh\barp_j+g^2\cosh\barp_k + \frac{1}{ \Phi_0^2}) \bch_i^{\rm hom}  - m_{ik}\,\bch_j^{\rm hom} -  m_{ij}\bch_k^{\rm hom}=0~,
\end{aligned}
\ee
again with $i\neq j\neq k$, and
\be
\begin{aligned}
m_{12}&=m_{21}= \frac{1}{\Phi_0^4}\le({\cal P}^1 {\cal P}^2 - {\cal P}^3 {\cal P}^4 \ri) e^{-\barp_1-\barp_2-\barp_3}~,\\
m_{13}&=m_{31}= \frac{1}{\Phi_0^4}\le({\cal P}^1 {\cal P}^3 - {\cal P}^2 {\cal P}^4\ri)  e^{-\barp_1-\barp_2-\barp_3}~,\\
m_{23}&=m_{32}= \frac{1}{\Phi_0^4}\le({\cal P}^2 {\cal P}^3 -{\cal P}^1 {\cal P}^4\ri)  e^{-\barp_1-\barp_2-\barp_3}~.
\end{aligned}
\ee

In Sec.\,\ref{sec:magneticnonBPS} and \ref{sec:magneticBPS}, we considered a simplification in which we further set three of the charges equal, leaving only two independent charges. Explicitly, in the notation we used for the non-BPS case, we set $P^2=P^3=P^1$, and $P^4$ is independent. After making this simplification, the cubic interactions are described by the effective action
\be \label{eq:cubicgeneral}
\begin{aligned}
    {\cal L}_{\rm int} & = \Y \partial_a\vec{\mathfrak{Z}}\cdot \partial^a\vec{\mathfrak{Z}} +a_1 \Y \left( \partial_a{\mathfrak{Z}_1} \partial^a{\mathfrak{Z}_1} + \partial_a{\mathfrak{Z}_2} \partial^a{\mathfrak{Z}_2} + \partial_a{\mathfrak{Z}_6} \partial^a{\mathfrak{Z}_6} \right) \\
    & \qquad + \frac{3}{4} a_1^2 \,\mathfrak{Z}_3  \partial^a \mathfrak{Z}_3 \partial_a \Y
    -\sum_{i=1}^6 \nu_i\, \Y \,\mathfrak{Z}_i\mathfrak{Z}_i~,
\end{aligned}
\ee
where the eigenstates in $\vec{\mathfrak{Z}}$ are orthogonal and appropriately normalized. We also have  
\be
\vec{a}= a_1(0,0,1,1,1,0)~, \qquad a_1\equiv 2\frac{1-e^{2\barp_1}}{1+3e^{2\barp_1}}~,
\ee
which parametrizes the inhomogeneous terms in \eqref{eq:inhomMagnetic}, and the cubic coefficients are
\be 
\begin{aligned}
    \nu_1&=\nu_2=a_1\le( \frac{(P^4)^2}{2\Phi_0^4}e^{-3\barp_1} -\frac{P^1P^4}{2\Phi_0^4}e^{-\barp_1}+g^2e^{-\barp_1}\ri) \\&\qquad\qquad+ \frac{3}{2}\frac{(P^4)^2}{\Phi_0^4}e^{-3\barp_1} +\frac{3}{2}\frac{P^1P^4}{\Phi_0^4}e^{-\barp_1}+g^2e^{-\barp_1}~, \\
    \nu_3&=-3 a_1^2  \le(\frac{(P^4)^2}{\Phi_0^4}e^{-3\barp_1}-g^2e^{-\barp_1}\ri) +2a_1 \le(\frac{(P^4)^2}{\Phi_0^4}e^{-3\barp_1}+6g^2\sinh(\barp_1)\ri)\\&\qquad\qquad+ 3\frac{(P^4)^2}{\Phi_0^4}e^{-3\barp_1} -g^2e^{-\barp_1}+2g^2e^{\barp_1}~,\\
    \nu_4&= \nu_5= -a_1 \le(\frac{(P^1)^2}{\Phi_0^4}e^{\barp_1}- g^2\sinh(\barp_1)\ri)+3\frac{(P^1)^2}{\Phi_0^4}e^{\barp_1}+g^2\cosh(\barp_1)~,\\
    \nu_6&=a_1\le(-\frac{9}{2}\frac{(P^1)^2}{\Phi_0^4}e^{\barp_1} + \frac{(P^4)^2}{2\Phi_0^4}e^{-3\barp_1} +\frac{P^1P^4}{\Phi_0^4}e^{-\barp_1}+g^2e^{-\barp_1}\ri) \\&\qquad\qquad +\frac{9}{2}\frac{(P^1)^2}{\Phi_0^4}e^{\barp_1} + \frac{3}{2}\frac{(P^4)^2}{\Phi_0^4}e^{-3\barp_1} -3\frac{P^1P^4}{\Phi_0^4}e^{-\barp_1}+g^2e^{-\barp_1}~.
\end{aligned}
\ee
The ungauged theory magnetic case is recovered by setting $a_1=0$ (i.e.\ by removing the inhomogeneous solution) and $g=0$. We stress again that these expressions apply for both BPS and non-BPS backgrounds.

In Sec.\,\ref{sec:gauged-cases}, we classify the corrections to the two-point functions of the fields $\mathfrak{Z}_i$ in terms of the parameter $\hat{D}$ defined in \eqref{eq:Dhat}. For the cubic interactions given in \eqref{eq:cubicgeneral}, the effective coupling constants are given by
\begin{equation}
    \begin{aligned}
        \lambda^{\rm eff}_1 &=\lambda^{\rm eff}_2 =   e^{-\barp_1}g^2+\frac{15}{2} e^{\barp_1}g^2-5 e^{-\barp_1}\frac{P^1P^4}{2\Phi_0^4} -\frac{3}{2\ell_2^2} \\
        &\qquad \quad + a_1\left(e^{-\barp_1}\, g^2+\frac{9}{2}e^{\barp_1}\, g^2 +e^{-\barp_1}\frac{P^1P^4}{2\Phi_0^4}-\frac{1}{2\ell_2^2}\right)~,\\
        \lambda^{\rm eff}_3 &=3 e^{-\barp_1}g^2+13 e^{\barp_1}g^2 -\frac{4}{\ell_2^2} +a_1\left(6 e^{-\barp_1}g^2 -\frac{2}{\ell_2^2}\right)-a_1^2\left(3 e^{-\barp_1}g^2+ 9 e^{\barp_1}g^2 -\frac{3}{2\ell_2^2}\right) ~,\\
        \lambda^{\rm eff}_4 &=\lambda^{\rm eff}_5=\frac{21}{2} e^{-\barp_1}g^2+\frac{11}{2} e^{\barp_1}g^2 -\frac{4}{\ell_2^2} -a_1\left(3\cosh\barp_1\, g^2 -\frac{1}{\ell_2^2}\right) ~,\\
        \lambda^{\rm eff}_6&= 16 e^{-\barp_1}g^2+15 e^{\barp_1}g^2+ 5 e^{-\barp_1}\frac{P^1P^4}{\Phi_0^4} -\frac{9}{\ell_2^2} \\
        &\qquad \quad -a_1\left(2e^{-\barp_1}\, g^2-3e^{\barp_1}\, g^2 -e^{-\barp_1}\frac{P^1P^4}{\Phi_0^4}-\frac{1}{\ell_2^2}\right) ~.
    \end{aligned}
\end{equation}

\renewcommand{\arraystretch}{1.4}

\section{Near-horizon behaviour of black holes in ungauged theory} \label{app:scalings}
In Sec.\,\ref{sec:linearUngauged} we discussed the eigenvectors of the linearized equations and their conformal dimensions. We showed that in the non-BPS case, the eigenvectors are linear combinations of the six scalar fields; some of them have conformal dimension $\Delta \neq 2$ and hence a different scaling near the horizon than the expected growth discussed in Sec.\,\ref{sec:nAtt}. In this appendix, we will give some details on how to take the extremal and near-horizon limits for the eigenstates in both the BPS and non-BPS branches, such that one finds the correct near-horizon response. 

To define the (ungauged, static) extremal limits, we will start from the black hole solution in \cite{Chow:2014cca}. This solution is parametrized in terms of one mass parameter $m_0$ and eight charge parameters ($\delta_I, \gamma^I)$. The metric is  
    \begin{equation}
        ds^2 = - \frac{R(r)}{W(r)} \dd t^2 + \frac{W(r)}{R(r)} \dd r^2 + W(r) (\dd \theta^2 + \sin^2\theta \dd \phi^2)~,
    \end{equation}
where $R(r)$ is a quadratic and $W^2(r)$ a quartic polynomial in $r$; these functions furthermore depend on the mass parameter $m_0$ and charge parameters $(\delta_I,\gamma_I)$. If we perform a dimensional reduction of this metric, it is clear from \eqref{eq:4d2d} that the dilaton is given by $\Phi (r) = \sqrt{W(r)}$. The precise definitions of $W(r)$ and $R(r)$, as well as those of the dilatons $\varphi_i$, axions $\chi_i$ and the electric and magnetic charges $(P^I, Q_I)$, can be found in \cite{Chow:2014cca}. Relevant to the discussion here is that at extremality and in the strict near-horizon limit $r\to 0$,  
    \begin{equation}
        R(r) \underset{r\to0}{=}   r^2~, \quad W(r) \underset{r\to0}{=} \Phi_0^2 = \ell_2^2~, 
    \end{equation}
such that we recover AdS$_2$ as the background metric; relative to the parametrization in \eqref{eq:poincare} we have $z=\ell_2^2/r$. For both the BPS and the non-BPS branch, the extremal limit consists of scaling
    \begin{equation}
        m_0 \sim m_0 \,\epsilon^2~, \quad \delta_I \sim \delta_I \epsilon^0~,
    \end{equation}
with $\epsilon \to 0$. This also implies that the horizon radius $r_h \sim m_0 \epsilon^2 \to 0$. The scaling of the $\gamma_I$ determines in which branch we land. We summarize the different possibilities in Table \ref{tab:extlimits}. 
\begin{table} 
\begin{center}
\begin{tabular}{|l|l|}
\hline
\textbf{Extremal limit} & \textbf{Attractor Solution} \\ \hline
$e^{\gamma_{\scriptscriptstyle I}} \sim e^{\gamma_{\scriptscriptstyle I}} \epsilon^{-1} $ or $e^{\gamma_{\scriptscriptstyle I}} \sim e^{\gamma_{\scriptscriptstyle I}} \epsilon $ & BPS: ~ ${\cal P}^1 = {\cal P}^2 = {\cal P}^3 = {\cal P}^4 ~, ~~ {\cal Q}_1 = {\cal Q}_2 = {\cal Q}_3 = {\cal Q}_4$   \\ \hline
$e^{\gamma_{\scriptscriptstyle{1,2}}} \sim e^{\gamma_{\scriptscriptstyle{1,2}}} \epsilon^{-1} ~\& ~ e^{\gamma_{\scriptscriptstyle{3,4}}} \sim e^{\gamma_{\scriptscriptstyle{3,4}}} \epsilon $ & BPS: ~ ${\cal P}^1 = {\cal P}^2 = - {\cal P}^3 = - {\cal P}^4~, ~~  {\cal Q}_1 = {\cal Q}_2 = - {\cal Q}_3 = - {\cal Q}_4$ \\ \hline
$e^{\gamma_{\scriptscriptstyle{1}}} \sim e^{\gamma_{\scriptscriptstyle{1}}} \epsilon^{-1} ~\& ~ e^{\gamma_{\scriptscriptstyle{i \neq 1}}} \sim e^{\gamma_{\scriptscriptstyle{i \neq 1}}} \epsilon $ & non-BPS \\ \hline
$e^{\gamma_{\scriptscriptstyle 1}} \sim e^{\gamma_{\scriptscriptstyle{1}}} \epsilon^{-1} ~\& ~ e^{\gamma_{\scriptscriptstyle{i \neq 1}}} \sim e^{\gamma_{\scriptscriptstyle{i \neq 1}}} \epsilon^0 $ & non-BPS \\ \hline
\end{tabular}
\caption{Different possible extremal limits, up to permutations. The second column contains its relation to solutions of the attractor equations in App.\,\ref{app:ads2} for $g=0$. In the non-BPS case the solutions are more complex and depend on the detailed configuration. }\label{tab:extlimits}
\end{center}
\end{table}
For the BPS branch, the choice of extremal limit also determines the number of relative minus signs. For both extremal limits that give the non-BPS branch, it is possible to tune the $\delta_I$ in such a way that one gets all or some of the ${\cal P}^I$ and ${\cal Q}_I$ charges equal (up to minus signs). An easy example for the second non-BPS extremal limit in Table \ref{tab:extlimits} is
    \begin{equation}
        \delta_1 = \delta_2 = \delta_3 \quad \Rightarrow \quad {\cal P}^1 = -{\cal P}^2=-{\cal P}^3=-{\cal P}^4~, ~ {\cal Q}_1 = {\cal Q}_2 = {\cal Q}_3 = - {\cal Q}_4~,
    \end{equation}
where $\delta_3 \neq 0$. These conditions on $\delta_I$ do not impose that any of the physical charges $P^I$ and $Q_I$ are equal. Note, however, that from this example it is clear that this non-BPS limit does not include the purely magnetic case (for which all $\delta_I = 0$). In general, both non-BPS limits assume that the charge parameters $(\delta_I, \gamma_I)$ are finite; for special cases where (some of) the charges vanish, one should first impose that, before taking an extremal limit.\footnote{In doing so, one will find that neither non-BPS limit in Table \ref{tab:extlimits} accommodates the purely electric case $\gamma_I = 0$. To get this, one has to switch the scalings of $\delta_I$ and $\gamma_I$  of the first non-BPS limit: take all $\gamma_I\sim \epsilon^0$ and impose the suitable scalings on $\delta_I$.} Finally, by choosing different numerical values for the $\delta_I$ and $\gamma_I$, one will end up in a non-BPS solution that does not comply with \eqref{eq:BPSequalPQs}; it can be checked numerically that the eigenvalues are still as in \eqref{eq:massNonBPS}. 

In Sec.\,\ref{sec:nAtt}, we discussed the linear response of the dilaton and axion fields away from their attractor values. At extremality ($\epsilon = 0$), the horizon radius is at $r_h = 0$, and we can expand the scalar fields around it to find in the BPS branch (for any extremal limit):
    \begin{equation}
        \begin{aligned}
            \varphi_i &= \barp_i + \varphi_i^{(1)} r + \cdots~, \\
            \chi_i &= \barc_i + \chi_i^{(1)} r + \cdots~,
        \end{aligned}
    \end{equation}
as predicted by the nAttractor mechanism \cite{Larsen:2018iou}. For a general non-BPS solution, this will still be true. However, for the non-BPS case, the single scalar fields are generally not eigenstates of the linearized equations of motion. Considering instead the near-horizon expansion of the eigenstates, we find again a linear response for the eigenvectors with $m^2 = \frac{2}{\ell_2^2}$, but for the eigenvectors with $m^2 = 0$ or $m^2 = \frac{6}{\ell_2^2}$ we find that the linear term cancels exactly. For some special cases that comply with \eqref{eq:BPSequalPQs}, this will already happen at the level of the dilatons and axions: for example, for the case discussed around \eqref{eq:nonBPSEV}, the linear terms of $\bch_1$ and $\bch_3$ vanish (as could have been expected from the eigenvector ${\mathfrak{Z}}_1$). 

To illustrate this cancellation let us consider the following numerical example for the first non-BPS limit in Table \ref{tab:extlimits}:
    \begin{equation}
        e^{\delta_1} = \frac{1}{2}~,~ e^{\delta_2} = 3~, ~e^{\delta_3} = 5~, ~e^{\delta_4} = 7~, ~ e^{\gamma_1} = 11 \epsilon^{-1}~, ~ e^{\gamma_2} = 13~, ~ e^{\gamma_3} = \frac{1}{11}~,~  e^{\gamma_4}=19~.
    \end{equation}
The eigenstates of the mass matrix are
    \begin{equation} \label{eq:numEV}
    \begin{aligned}
      m^2 \ell_2^2 &= 0~: ~\,\quad
      \begin{array}{l}
      {\mathfrak{Z}}_1 = -0.1826 \,\bph_1 + 0.928779\,\bph_3 + 1.35251\,\bch_1 + \bch_3 ~, \\
       {\mathfrak{Z}}_2 = -0.133794\, \bph_1 + 1.26839 \cdot 10^{-5} \,\bph_2 + 0.991009 \,\bch_1 + \bch_2~,
      \end{array}
 \\[0.6em]
      m^2 \ell_2^2 &= 2~:~\, \quad  
      \begin{array}{l}
      {\mathfrak{Z}}_3 = 7.40696 \,\bph_1 +  \bch_1~, \\  
      {\mathfrak{Z}}_4 = -7.88398 \cdot 10^4\, \bph_2 +  \bch_2~, \\
      {\mathfrak{Z}}_5 =   -1.07668\,\bph_3 + \bch_3~,     
      \end{array}
      \\[0.6em]
      m^2\ell_2^2 &=6~:~\, \quad 
      \begin{array}{l}
      {\mathfrak{Z}}_6 = 0.1826\, \bph_1 + 1.73108 \cdot 10^{-5} \,\bph_2 + 0.928779 \,\bph_3 \\
      \qquad - 1.35251\, \bch_1 + 1.36478 \, \bch_2 +  \bch_3~, 
      \end{array}
    \end{aligned}        
    \end{equation}
This is what our analysis predicts for the eigenstates. On the other hand, if we take the corresponding on-shell solution in \cite{Chow:2014cca} with these same charges, and upon taking the near-horizon limit $r\to 0$, the moduli all have linear responses: 
    \begin{equation}
    \begin{split}
        \bch_1 &= \barc_{1,0} - 0.122629\, r + O(r^2)~ \\
        \bch_2 &= \barc_{2,0} - 3.1277 \cdot 10^{-7} \, r + O(r^2)~ \\
        \bch_3 &= \barc_{3,0} + 8.49969\, r + O(r^2)~ 
    \end{split}
    \quad ~, \quad 
    \begin{split}
        \bph_1 &= \barp_{1,0} -2.30464\, r + O(r^2)~ \\
        \bph_2 &= \barp_{2,0} + 5.59775\, r + O(r^2)~ \\
        \bph_3 &= \barp_{3,0} - 1.70457 \,r + O(r^2)~
    \end{split}
    \quad ~,
    \end{equation}
but if we consider the near-horizon linear combinations dictated by  the eigenvectors \eqref{eq:numEV} we find
    \begin{equation}
    \begin{aligned}
      m^2 \ell_2^2 &= 0~: ~\,\quad {\mathfrak{Z}}_1 = -2.23152 + O(r^2) ~, \quad {\mathfrak{Z}}_2 = 6332.26 + O(r^2)~, \\
      m^2 \ell_2^2 &= 2~:~\, \quad {\mathfrak{Z}}_3 = 6.18617 - 17.3815\,r + O(r^2)~, \quad {\mathfrak{Z}}_4 = -784443  -441326\,r + O(r^2)~, \\
      &\quad \quad \quad \quad \,\,\, {\mathfrak{Z}}_5 =   2.26974 + 3.42844\,r + O(r^2)~,\\
      m^2\ell_2^2 &=6~:~\, \quad {\mathfrak{Z}}_6 = 8643.32 + O(r^2)~,  
    \end{aligned}           
    \end{equation}
Thus, indeed, the eigenvectors with $m^2\ell_2^2 = 2$ still have a linear response, as predicted, but for the eigenvectors with $m^2\ell_2^2 = \{0, 6\}$ this linear term vanishes in accordance with \eqref{eq:fraknonBPS}. 

It also worth mentioning that the  non-BPS eigenstates with $m^2\ell_2^2 = \{0, 6\}$ is easily overlooked in simple cases. For example, in the purely magnetic or purely electric non-BPS case, the eigenvectors are
    \begin{equation}
    \begin{aligned}
      m^2 \ell_2^2&= 0~: \quad {\mathfrak{Z}}_1 = \bch_1 + \bch_2 ~, \quad {\mathfrak{Z}}_2 = \bch_1 + \bch_3 ~, \\
      m^2 \ell_2^2 &= 2~: \quad {\mathfrak{Z}}_3 = \bph_1~, \quad {\mathfrak{Z}}_4 = \bph_2~, \quad {\mathfrak{Z}}_5 =  \bph_3 ~,\\
      m^2 \ell_2^2 &= 6~: \quad {\mathfrak{Z}}_6 = - \bch_1 + \bch_2 + \bch_3~,
    \end{aligned}        
    \end{equation}
but for these cases the axions actually vanish for the corresponding black hole, such that we are left only with the eigenvectors $\bph_i$, which have the usual linear response. 

\bibliographystyle{JHEP-2}
\bibliography{all}

\end{document}